
\input harvmac.tex
\noblackbox
\lref\Smir{F. Smirnov, ``Form factors in completely
integrable models of quantum field theory'', World Scientific and
references therein.}
\lref\Kar{M. Karowski, P. Weisz, Nucl. Phys. B{\bf 139} (1978) 455.}
\lref\W{K. M. Watson, Phys. Rev. B{\bf 95} (1954) 228.}
 \lref\CarMus{J. Cardy, G. Mussardo, Nucl. Phys. B{\bf 410} (1993),
451.}
\lref\Musmisc{G. Delfino, G. Mussardo, P. Simonetti, Phys. Rev. D{\bf
51}
(1995), 6620.}
\lref\ZZ{A, Zamolodchikov, Al. Zamolodchikov, Ann. Phys. {\bf 120}
(1979) 253.}
\lref\Bar{R. Z. Bariev, Teor. Mat. Fiz. {\bf 40} (1979), 40; Teor.
Mat.
Fiz. {\bf 42} (1980),
262; Teor. Mat. Fiz. {\bf 77} (1988), 1090.}
\lref\Korbook{The review V. E. Korepin, N.M Bogoliubov, A.G. Izergin,
``Quantum
Inverse Scattering Method and Correlation Functions", Cambridge
University Press (1993) and references therein.}
\lref\GZ{S. Ghoshal, A. Zamolodchikov, Int. J. Phys. A {\bf 9},
3841 (1994).}
\lref\Hecht{R. Hecht, Phys. Rev. B {\bf 138} (1967) 557.}
\lref\IS{C. Itzykson, H. Saleur, J. Stat. Phys. {\bf 48} (1987) 449.}
\lref\mccoy{T.T. Wu,
B.M. McCoy, C.A. Tracy and E. Barouch, Phys. Rev. B {\bf 13}, 316
(1976);
B.M. McCoy and T.T. Wu, Phys. Rev. Lett. {\bf 45}, 675 (1980);
J.H.H. Perk, Phys. Lett. {\bf 79A}, 1, 3 (1980); B.M. McCoy, J.H.H.
Perk,
and T.T. Wu, Phys. Rev. Lett. {\bf 46}, 757 (1981).}
\lref\perk{J.H.H.
Perk, H.W. Capel, G.R.W. Quispel, and F.W. Nijhoff, Physica
 {\bf 123A}, 1 (1984).}

\def\zbar{{\bar{z}}}
\def\>{\rangle}
\def\<{\langle}

\Title{CLNS 96/1410, USC-96-013, cond-mat/9606104}
{\vbox{
\centerline{Finite temperature correlations in the}
 \vskip 4pt
\centerline{one-dimensional quantum Ising model.}
}}
\centerline{A. Leclair$^1$, F. Lesage$^2$, S. Sachdev$^3$, H.
Saleur$^2$}
\smallskip
\centerline{$^1$Newman Laboratory, Cornell University}
\centerline{Ithaca, NY 14853}
\smallskip
\centerline{$^2$Department of Physics}
\centerline{University of Southern California}
\centerline{Los Angeles, CA 90089-0484}
\smallskip
\centerline{$^3$ Department of Physics, Yale University,}
\centerline{P.O. Box 208120, New Haven CT 06520-8120}

\vskip .3in

Abstract: We extend the form-factors approach to the quantum Ising
model
at finite temperature. The two point function of the energy is
obtained in
closed form,
while the two point function of the spin is written as a Fredholm
determinant.
Using the approach of \Korbook, we obtain, starting directly
from the continuum formulation,  a set of  six differential
equations
satisfied by this two point function. Four of these equations
involve only spacetime derivatives, of which three
are equivalent  to the  equations
obtained earlier in \mccoy,\perk. In addition, we obtain two new
equations
involving a temperature derivative. Some of these results are
generalized
to the Ising model on the half line with a magnetic field
at the origin.

\Date{04/96}

%
%
%

%
%

\def\tilde{\widetilde}
\def\bar{\overline}
\def\hat{\widehat}
\def\*{\star}
\def\[{\left[}
\def\]{\right]}
\def\({\left(}		
\def\){\right)}

%
%
\def\zb{{\bar{z} }}
\def\frac#1#2{{#1 \over #2}}
\def\inv#1{{1 \over #1}}

\def\d{\partial}

\def\2pi{\hbox{$2\pi i$}}

\def\dsl{\raise.15ex\hbox{/}\kern-.57em\partial}
\def\Dsl{\,\raise.15ex\hbox{/}\mkern-.13.5mu D}
%
%
\def\th{\theta}

\def\ep{\epsilon}

%
%
\def\CA{{\cal A}}	\def\CB{{\cal B}}

		\def\CO{{\cal O}}

\def\2pi{\hbox{$2\pi i$}}

\def\dsl{\raise.15ex\hbox{/}\kern-.57em\partial}
\def\Dsl{\,\raise.15ex\hbox{/}\mkern-.13.5mu D}
%
%
%
\font\numbers=cmss12
\font\upright=cmu10 scaled\magstep1
\def\stroke{\vrule height8pt width0.4pt depth-0.1pt}
\def\topfleck{\vrule height8pt width0.5pt depth-5.9pt}
\def\botfleck{\vrule height2pt width0.5pt depth0.1pt}
\def\Zmath{\vcenter{\hbox{\numbers\rlap{\rlap{Z}\kern
0.8pt\topfleck}\kern 2.2pt
                   \rlap Z\kern 6pt\botfleck\kern 1pt}}}
\def\Qmath{\vcenter{\hbox{\upright\rlap{\rlap{Q}\kern
                   3.8pt\stroke}\phantom{Q}}}}
\def\Nmath{\vcenter{\hbox{\upright\rlap{I}\kern 1.7pt N}}}
\def\Cmath{\vcenter{\hbox{\upright\rlap{\rlap{C}\kern
                   3.8pt\stroke}\phantom{C}}}}
\def\Rmath{\vcenter{\hbox{\upright\rlap{I}\kern 1.7pt R}}}
\def\Z{\ifmmode\Zmath\else$\Zmath$\fi}
\def\Q{\ifmmode\Qmath\else$\Qmath$\fi}
\def\N{\ifmmode\Nmath\else$\Nmath$\fi}
\def\C{\ifmmode\Cmath\else$\Cmath$\fi}
\def\R{\ifmmode\Rmath\else$\Rmath$\fi}

\newsec{Introduction}

The form-factors approach has been a surprisingly
efficient tool for computing correlation functions
in $1+1$ quantum integrable theories at zero temperature ($T$)
(see eg. \W,\Kar,\Smir,\CarMus,\Musmisc).
Recall that a quantum integrable theory  in $1+1$   is conveniently
described
in terms of particles (massive or massless), with a  factorized
scattering
encoded in  the  $S$ matrix (solution of the Yang-Baxter equation).
Multiparticle
states provide a natural basis for the space of states of the theory
\ZZ,\Smir.
The form-factors are the matrix elements of the physical operators
in this basis. In most theories, they can be obtained by  solving
a set of axioms generalizing Watson's equation \W. Correlation
functions are then expressed as  infinite series of products of these
 matrix elements. While
exact, re-summed expressions are hard to obtain, the approach is
usually
very satisfactory from a practical  point of view:  the series
converge rather quickly \Smir,\CarMus, allowing in particular
potential
comparison with numerical simulations.

So far, only integrable quantum theories at $T=0$ have been
considered
(equivalently,
statistical mechanics models on the plane). For the purpose of
comparison with experiments however, it is crucial to obtain
results at finite temperature (which corresponds in statistical
mechanics to
models on a cylinder). This seems to be a rather challenging
subject. In \ref\LS{F. Lesage, H. Saleur, ``Exact results
for correlators in quantum impurity problems
with an external field'', preprint USC-96-09, to appear.}, a related
problem
has been considered, that of $T=0$ but a non-vanishing chemical
potential. The
 main
difficulty encountered there was that the non vanishing potential
changes the nature of the ground state, and to use the form-factor
approach one has to compute matrix elements
of operators between states with a  finite
number of particles per unit length. This is rather delicate; in
fact,
simply computing the scalar product of such states has been
a major achievement in the study of integrable systems \ref\Gaudin{M.
Gaudin,
``La fonction d'onde de Bethe'', Masson (1983)},
\Korbook.

In the  present paper, we present a first step
towards finite temperature results by treating in
detail the quantum
Ising model in one dimension ($1d$). Many of the difficulties
expected in general integrable systems, disappear in that
case due to the simplicity of the $S$ matrix, $S=-1$. Some of them
remain however, especially for the spin operator which is non
local in terms of fermions.

The $1d$ quantum Ising model also presents physical interest on its
own.
It presents  a quantum-critical point
between an ordered ground state with a broken symmetry and a quantum
paramagnetic phase, and is
 the
simplest system for which finite temperature crossovers near quantum
critical
points can be exactly computed.  The
crossovers are   non-trivial and most of their
qualitative features  do  generalize to higher dimensions. These
exact results are therefore an important testing ground for our
current
understanding of these crossovers \ref\hertz{J.A. Hertz, Phys. Rev. B
{\bf 14}, 525 (1976).} \ref\chn{S. Chakravarty, B.I. Halperin, and
D.R. Nelson,
Phys. Rev. B {\bf 39}, 2344 (1989).} \ref\csy{S. Sachdev and J. Ye,
Phys.
Rev. Lett. {\bf 69}, 2411 (1992); A.V. Chubukov and S. Sachdev, Phys.
Rev.
Lett.{\bf 71}, 169 (1993); A.V. Chubukov,  S. Sachdev and J. Ye,
Phys.  Rev. B
{\bf 49}, 11919 (1994).}. Indeed, as was recently noted
\ref\subir{S. Sachdev, Nucl. Phys. B {\bf 464}, 576 (1996); K. Damle
and S.
Sachdev, Phys. Rev. Lett. {\bf 76}, 4412 (1996).}, there are a number
of striking
similarities
between the crossovers of the $d=1$ quantum Ising model and those of
the $d=2$
$O(3)$ quantum rotor model (the latter is a useful model of
anti-ferromagnetic spin
fluctuations in the cuprate superconductors \chn,\csy.) Finally, it
is not
unreasonable to hope that detailed low temperature studies of
experimental
realizations of the $d=1$ Ising model will emerge in the future.

The paper is organized as follows. In sections 2 and 3
we develop a method to handle correlators at finite temperature. This
method focuses onto excitations over the ``thermal ground state''
(rather
than the usual vacuum of the theory), and seems the most suitable
to generalizations to truly interacting theories \LS. The method is
explained
in more detail in  Section 2,
where we  derive form-factors expansions for the $d=1$ quantum Ising
correlators
at finite temperature. The two point function of the energy is
explicitly
computed, and  a Fredholm determinant
expression is obtained for the two point function of the spin. In
Section 3,
the analysis
is extended to the model on the half line with a magnetic field at the
boundary. Again, the
one point function of the energy is computed, and a Fredholm
determinant
 expression
for the one point function of the spin is obtained.
In section 4, we return to the bulk problem, and obtain differential
equations
satisfied by the two-point spin correlators. Earlier work \mccoy\
 had obtained
differential equations for the
$d=2$ classical Ising model in the infinite plane (equivalent to the
$d=1$ quantum
Ising model at
$T=0$), and precisely the same equations were later shown in
\perk\
to also apply to the $d=1$ quantum Ising model at non-zero $T$. Among
our results
is a set of similar equations which are consistent with, but slightly
stronger
than, those of \perk, and obtained by a very different method. In
addition, we will
obtain entirely new equations which have the novel feature of
involving
derivatives with respect to temperature, and which are therefore
special to the
non-zero $T$ case. Differential equations for correlators with
temperature
derivatives were obtained earlier in \Korbook\ for the
non-relativistic dilute
Bose gas, and the methodology of Section 4 is similar. In the
conclusion, we comment on the differential equations
that should be satisfied by the one point function of the spin
with a boundary. We also put our results for the two-point spin
correlator
in the bulk in a more physical perspective.

\newsec{The form-factor approach to correlators of the  Ising model
at finite $T$.}

\subsec{Generalities}

We describe the quantum Ising model in 1d in terms
of massive fermions. The associated creation and
annihilation operators have the algebra:
\eqn\fadzamo{\eqalign{
A(\beta_1)A(\beta_2)&=-A(\beta_2)A(\beta_1)\cr
A^\dagger(\beta_1)A^\dagger(\beta_2)&=
-A^\dagger(\beta_2)
A^\dagger(\beta_1)  \cr
A(\beta_1)A^\dagger(\beta_2)&=-
A^\dagger(\beta_2) A(\beta_1)+
2\pi \delta(\beta_1-\beta_2).
}}
In these expressions, the $\beta_i$'s are the usual rapidity
variables parameterizing momentum and energy as
$p(\beta)=m\sinh\beta,e(\beta)=m\cosh\beta$.

Let us now consider the question of evaluating the  two point
function
of some operator ${\cal O}$ in the Ising model at finite temperature
$T$.  The space coordinate is $x$,  $-\infty< x < \infty$,
and we call  $t$  the time coordinate.
Formally one has:
\eqn\twopoint{<{\cal O}(x,t){\cal O}(0,0)>={1\over
Z}\sum_{\psi,\psi'}e^{-
E_{\psi} / T}
\  {<\psi|{\cal O}(x,t)|\psi'><\psi'|{\cal
O}(0,0)|\psi>\over <\psi|\psi> <\psi'|\psi'> },}
where the states $|\psi>$  are multiparticle
states,  the matrix elements of ${\cal O}$ are the so-called
form-factors, and $Z$ is the partition function which formally
corresponds to $Z = \sum_\psi e^{-E_\psi /T} $.

There are divergences in the expression \twopoint\ that are a result
of working with multiparticle states that are defined in infinite
volume (the $x$-direction).  These infinite volume divergences
manifest themselves in the norms of the states, e.g.
$<\beta | \beta > =2\pi \delta(0)$.  These can be regulated by
letting
the cylinder have a finite length $L$ and letting $L\to \infty$.
The $\delta (0)$ are then regularized to $ {mL\over 2\pi} \cosh
\beta$.
With this explicit regularization, it is natural to consider
a thermodynamic approach to the expression \twopoint, which we
describe
in the remainder of this section.
In this free field situation, it is possible
to deal with the expression \twopoint\ directly, and this
provides a check on the thermodynamic approach; this
is presented in Appendix A.

\subsec{Thermodynamic approach}

If we regularize
the norms as explained above,  the sum over intermediate states
is  a discrete sum over states with allowed momenta. On
general
grounds, we expect the partition function to go like $\exp ( L f)$
where
$f$ is the dimensionless free energy per unit length.
That $Z$ behaves exponentially
means that
the sum is dominated by terms where the energy goes like $L$, the
size of the system. In other words, the sum is dominated by
multiparticle states that have a number of particles proportional
to the size of the system $L$. These states we call
{\bf macroscopic}, and for our purpose they will be  completely
 characterized by
non-vanishing densities of particles (per unit length) as $L\to
\infty$.
This situation is to be contrasted with  computations at $T=0$, where
one
 considers only microscopic  states,  that have a finite
 number of particles as
$L\to \infty$.

Since microscopic states give a negligible contribution to \twopoint\
for $T\neq 0$, we can
replace  $\sum_{\psi}$ by a functional integral~:
\eqn\repl{\sum_{\psi}\to \int [d\rho],}
where $\rho$ denotes the density of particles
per unit length (the exact measure
to put in \repl\ deserves some discussion, but it will not matter in
what
follows). The denominator
and numerator become respectively
\eqn\systm{\eqalign{Z&=\int [d\rho] e^{-\beta \{
E([\rho])+S([\rho])\} }\cr
\hbox{Num }&=\int [d\rho] e^{-\beta \{
E([\rho])+S([\rho]) \}}
\sum_{\psi'}{
<[\rho]|{\cal O}(x,t)|\psi'><\psi'|{\cal O}(0,0)|[\rho]>\over
<\psi'|\psi'>} .\cr}}
Here $E[\rho]=m L\int d\theta \rho(\theta) \cosh\theta$, the entropy
follows from Stirling's formula $S=L\int d\theta [(\rho+\rho^h)\ln
(\rho+\rho^h)
-\rho\ln \rho-\rho^h\ln\rho^h]$, where $\rho^h$ is the density of
holes, $\rho$ the density of particles, with
$\rho(\beta)+\rho^h(\beta)=
{1\over 2\pi}m\cosh\beta$.
The point
now is that the argument of the numerator and the denominator \systm\
differ only by the correlator, which is of order one. In the limit
$L\to \infty$, both are dominated by the same saddle point,
characterized by
$\rho/\rho^h=\exp(-m\cosh\beta/T)$. From now on we reserve the
notation
$\rho$ for the saddle point density. Then the exponential
contributions  of the numerator and denominator
cancel out, and we get the simple result (a similar result appears in
a different context in \Korbook)~:
\eqn\simpres{<{\cal O}(x,t){\cal O}(0,0>={1\over
<0_T|0_T>}\sum_{\psi'} {<0_T|{\cal O}(x,t)|\psi'>
<\psi'|{\cal O}(0,0)|0_T>\over <\psi'|\psi'>}, }
where we denote by $|0_T>$ {\bf any} multiparticle state that is
characterized by the macroscopic density $[\rho]$ (which one does not
matter). We will refer to this state, a bit incorrectly,  as the
``thermal ground state''.

Now, the remaining task
is to evaluate the matrix elements.  There are several ways to
proceed,
 - here, we present
what seems the most physical approach, and probably the most suitable
to generalizations.

\subsec{Excitations over the thermal ground state}

The idea is not to use the usual form-factors, which correspond to
excitations
above the vacuum $|0>$, but rather to introduce new
form-factors
appropriate for the excitations
above $|0_T>$.  Consider therefore
a multiparticle state that realizes the density $\rho$. The
excitations
above this state  are of two kinds: one can add a particle,
or one can create a hole. We will take into
account these two possibilities by introducing two types of
``thermal'' particles, characterized by a label $\epsilon=\pm 1$.
Observe
that the sets of rapidities where a thermal particle with
$\epsilon=1$
and $\epsilon=-1$ can be created are separate. The
creation/annihilation
operators of thermal particles with the same $\epsilon$ satisfy
\fadzamo\
while operators with different $\epsilon$ anticommute. The
excitations
therefore still have factorized scattering with $S=-1$. We now define
the form-factors of an operator ${\cal O}$ as
%
\eqn\newffdef{F(\beta_1,\beta_2,...,\beta_n)_{\epsilon_1,...,\epsilon_ 
n}\equiv
{1\over \sqrt{<0_T|0_T>}}
{}^{\epsilon_n,\ldots,\epsilon_1}<\beta_n\ldots\beta_1|{\cal
O}|0_T>,}
where the subscript $T$ is implicit
for multiparticle states. These form-factors have to satisfy
relations that are very similar
to the ones for excitations over the ground state. One has
\eqn\axiomi{
F(\beta_1,\beta_2,...,\beta_n)_{\epsilon_1,...,\epsilon_n}=
-F(\beta_2,\beta_1,...,\beta_n)_{\epsilon_2,\epsilon_1,...,
\epsilon_n},
}
from the fact that  $S=-1$.  CPT invariance
 gives~:
\eqn\axiomii{
F(\beta_1,\beta_2,...,\beta_n+2\pi i)_{\epsilon_1,...,\epsilon_n}=
F(\beta_n,\beta_1,...,\beta_{n-1})_{\epsilon_n,\epsilon_1,...,
\epsilon_{n-1}}.
}
The third axiom is slightly more complicated. First, one expects as
usual an
annihilation pole at $\beta_i=\beta_j-i\pi$ when
$\epsilon_i=\epsilon_j$.
In addition, remembering the origin of $|0_T >$,
and the fact that  a hole and a particle can annihilate each
other, there should also be a pole at  $\beta_i=\beta_j$ when
 $\epsilon_i=-\epsilon_j$.
This leads to
\eqn\residues{\eqalign{
{\rm Res}_{\beta_1=\beta_2-i\pi}&F^{(n)}(\beta_1,\beta_2,...,
\beta_n)_{\epsilon_1,\epsilon_1,...\epsilon_n}=F^{(n-2)}
(\beta_3,...,\beta_n)_{\epsilon_3,...,\epsilon_n}
\cr
{\rm Res}_{\beta_1=\beta_2} & F^{(n)}(\beta_1,\beta_2,...,
\beta_n)_{\epsilon_1,-\epsilon_1,...\epsilon_n}=F^{(n-2)}
(\beta_3,...,\beta_n)_{\epsilon_3,...,\epsilon_n}
}}

There are two minimal  solutions to these equations. The one
corresponding
to the energy operator is
\eqn\enedens{\eqalign{
F(\beta_1,\beta_2)_{++}&=m \sinh ({\beta_1-\beta_2\over 2}) \cr
F(\beta_1,\beta_2)_{+-}&=im\cosh ({\beta_1-\beta_2\over 2}) \cr
F(\beta_1,\beta_2)_{-+}&=-im\cosh ({\beta_1-\beta_2\over 2}) \cr
F(\beta_1,\beta_2)_{--}&=m\sinh ({\beta_1-\beta_2\over 2})
}}
The solution  corresponding to the spin $\sigma$  and disorder
operators
$\mu$ is
\eqn\solution{
F(\beta_1,...,\beta_n)_{\epsilon_1,...,\epsilon_n}=  i^{n/2}
\prod_{i<j} \tanh \left( {\beta_i-\beta_j\over 2}\right)^{
\epsilon_i \epsilon_j}  ,}
where $n$ is even for $\sigma$, with $<0_T|\sigma|0_T>/<0_T|0_T>=1$,
and $n$ is odd for $\mu$. Note that we have chosen to work
in the ordered phase here, and  that
the form-factors approach requires us  to consider the subtracted energy
operator (of vanishing one-point function), while the spin is not
subtracted.
In \enedens\ as well as \solution, the
overall normalization is largely arbitrary.

In this free field situation, the
particle and hole form factors are deduced from
the  usual ones for particles only by shifting the
hole rapidities
by $i\pi$. This hints at a more direct way of obtaining expressions
for the correlators, which is explained in appendix A.

Knowing the form-factors, the two point function of the operator
${\cal O}$
can finally be computed. A little subtlety appears in the sum over
intermediate
states. While multi-thermal particle states are normalized as usual,
their allowed rapidities are distributed according to the filling
fractions
\eqn\filling{
f_\epsilon(\beta)={1\over 1+e^{-\epsilon m\cosh\beta/T}}.
}
One finds therefore
\eqn\generes{\eqalign{<{\cal O}(x,t){\cal O}(0,0)>=&\sum_{n=0}^\infty
{1\over  n!}\sum_{\epsilon_i}
\int_{-\infty}^\infty \prod_{i=1}^n \left\{{d\beta_i\over 2\pi}
 f_{\epsilon_i}(\beta_i)
e^{-i\epsilon_i [mt \cosh \beta_i-mx\sinh\beta_i]} \right\}\cr
&~~~~~~~~~~~~~~
\times|F(\beta_1,\ldots,\beta_n)_{\epsilon_1,\ldots,\epsilon_n}|^2.\cr 
}}

It is not difficult to verify that the above expression satisfies
the Martin-Schwinger-Kubo relation:
\eqn\marts{
<\CO (x, t-i/T) \CO (0) > =
<\CO (0) \CO(x,t) > . }

\subsec{The energy correlations}

As a check, it  is useful to consider the
conformal limit of the energy two-point function, which can be
compared
 with standard results. We obtain the conformal correlator
by the limit
\eqn\massle{\eqalign{
\beta&=\pm\beta_0\pm\theta , \ \ \beta_0\rightarrow \infty
\cr
m\to 0&,{m\over 2} e^{\beta_0}\to \mu.
}}
Here $\mu$ is a parameter adjusting
the normalization. The spectrum separates into left and right
moving particles
with $e=\mu e^{\theta}=\pm p$.
In this limit, only
the left-right and right-left form factors will be non zero.
They are simply~:
\eqn\massff{\eqalign{
F(\theta_1,\theta_2)^{RL}_{++}&=\mu e^{(\theta_1+\theta_2)/ 2} \cr
F(\theta_1,\theta_2)^{RL}_{+-}&=i\mu e^{(\theta_1+\theta_2)/2}\cr
F(\theta_1,\theta_2)^{RL}_{-+}&=-i\mu e^{(\theta_1+\theta_2)/2}\cr
F(\theta_1,\theta_2)^{RL}_{--}&=\mu e^{(\theta_1+\theta_2)/2}.\cr
}}
The two point function of the energy factorizes
as  the  product of a right and a left moving part.
For
 the right moving
part we find~:
\eqn\holo{\eqalign{
<e(x,t)e(0,0)>_R&=\mu^2\int_{-\infty}^\infty {d\theta\over 2\pi}
[f_+(\theta)e^{-2z_R\mu e^\theta}
+f_-(\theta)e^{2z_R\mu e^\theta}]
e^\theta  \cr
&=\mu\int_0^\infty {du\over 2\pi} \left[{e^{-2uz_R}\over 1+e^{-u/T}}+
{e^{2uz_R}\over 1+e^{u/T}} \right]\cr
&={\mu T\over
2\sin(2\pi
T z_R)},
}}
where we have set $z_R=(t-ix)/2$ (and similarly $z_L=(t+ix)/2$). This
is the
well known   conformal result
for the fermions correlator, and multiplying by its conjugate
gives of course the  correct energy correlator.

In the massive case, the holomorphic and anti-holomorphic
parts will not be decoupled and the solution will be more
involved. Equation \generes\ gives,
 restricting to $x=0$ for simplicity~:
\eqn\maener{\eqalign{
&~~~~~~~~~~~~~~~~~~~~~{(2\pi)^2\over m^2}<e(0,t)e(0,0)>=\cr
&\int_0^\infty du_1 du_2
{e^{imt(u_1+1/u_1)/2}
\over 1+e^{m (u_1+1/u_1)/2T}}
{e^{imt(u_2+1/u_2)/2}\over 1+e^{m (u_2+1/u_2)/2T}} \left({1\over
u_1}-
{1\over u_2}\right)^2\cr &+ \left({1\over
u_1}-
{1\over u_2}\right)^2 +
{e^{imt(u_1+1/u_1)/2}\over 1+e^{m (u_1+1/u_1)/2T}}
{e^{-imt(u_2+1/u_2)/2}\over 1+e^{-m (u_2+1/u_2)/2T}} \left({1\over
u_1}+
{1\over u_2}\right)^2\cr & +
{e^{-imt(u_1+1/u_1)/2}\over 1+e^{-m (u_1+1/u_1)/2T}}
{e^{imt(u_2+1/u_2)/2}\over 1+e^{m (u_2+1/u_2)/2T}} \left({1\over
u_1}+
{1\over u_2}\right)^2\cr & +
{e^{-imt(u_1+1/u_1)/2}\over 1+e^{-m (u_1+1/u_1)/2T}}
{e^{-imt(u_1+1/u_2)/2}\over 1+e^{-m (u_1+1/u_1)/2T}} \left({1\over
u_1}-
{1\over u_2}\right)^2,
}}
where we have introduced the variable $u=e^\beta$.
The integrated result is given by~:
\eqn\inresu{\eqalign{
{(2\pi)^2\over m^2}<e(0,t)e(0,0)>=~~~~~~~~~~~~~~~~~~~~\cr
 \left\{\sum_{n=0}^\infty (-1)^n
\left[ K_1(imt+nm/T)
+K_1(nm/T+m/T-imt)\right]\right\}^2\cr -
\left\{ \sum_{n=0}^\infty (-1)^n\left[
K_0(imt+nm/T)-K_0(nm/T+m/T-imt)\right]\right\}^2.\cr
}}
Here $K$ are Bessel functions of the second kind.
It reproduces the conformal  result in the limit $m\to 0$.  In the
limit
$T\to 0$ it also reproduces the result of \Hecht.

\subsec{The spin-disorder correlations}

The conformal spin correlator cannot be naively recovered
by the limit \massle, the integrals being infrared divergent. In
fact,
the understanding of the spin correlator is considerably more
involved,
and will be the subject of most of this paper.

Let us continue to Euclidean space
 space $t\to -it$, and define
$z = (t-ix)/2 $, $\zbar = (t+ix)/2$.
Define
\eqn\eIIIv{
\tau_{\pm} = \< \sigma (z, \zb ) \sigma (0) \> \pm \< \mu (z, \zb )
\mu (0)
\> . }
It is convenient to express everything in terms of the  variable $u$:
\eqn\eIIIii{ u = e^\th .}
Then, the formula \generes\ leads to
\eqn\eIIIvi{\eqalign{
\tau_{\pm} =&
\sum_{N=0}^\infty
\frac{(\pm 1)^N}{N! } \sum_{\ep_1, \ep_2 ,..=\pm 1}
\int_0^\infty
\frac{du_1}{2\pi u_1}
\cdots \frac{du_N}{2\pi u_N}
\[ \prod_{i=1}^N f_{\ep_i} (u_i) e^{-\ep_i (mzu_i +   m \zb /u_i ) }
\]\cr
&~~~~~~~~~~~~~~~~~~~~~~~~~~~~~\times\[ \prod_{i<j} \( \frac{u_i - u_j
}{u_i + u_j} \)^{2 \ep_i \ep_j }  \]
,\cr}}
where it should be clear from the context that the thermal filling
factors
have changed their definition from \filling\ to one appropriate for
an energy
expressed in terms of the $u$ variable
\eqn\fillingu{
f_{\epsilon} (u) = \frac{1}{1 + e^{-\epsilon m (u+1/u)/2 T}}.}
At $T=0$ \eIIIvi\ reduces to  well known expressions (see {\it
e.g.\/}
\ref\YZ{V. P. Yurov,
 Al. B. Zamolodchikov, Int. J. Mod. Phys. A {\bf 6} (1991) 33419.},
\ref\bernard{
O. Babelon and D. Bernard, Physics Letters. B {\bf 288},
113 (1992).}).
\def\Det{{ \rm Det}}
\def\det{{ \rm det}}

The poles in \eIIIvi,
which  are due to particle  hole annihilation,
are now clearly seen as
 an artifact of the
 continuum limit.  Consider the case where  $L$ is finite
 and the particles
 have quantized momenta.  Then,
 a particle and a hole cannot have the same
 rapidity: a state is either filled or empty.  When replacing
 discrete sums by integrals, this exclusion disappeared,
 and has to be reinstated by hand. The prescription will be
 to move slightly one of the rapidities off the real axis.
 This is what will be understood in the following.

The functions $\tau_{\pm}$ can be expressed as a Fredholm determinant
of
a $2\times 2$ matrix of integral operators
\eqn\eIIIvic{
\tau_{\pm} = \Det ( 1 \pm {\bf W} ), }
where
\eqn\eIIIvib{
{\bf W} (u,v)  = \left(\matrix{W_{++} (u,v)  & W_{+-} (u,v)
\cr W_{-+} (u,v)  & W_{--} (u,v)  \cr }\right)
. }
The two dimensional vector space structure arises from the
$\ep$ indices distinguishing particles and holes.
Extending the Fredholm theory to this case, one
has by definition:
\eqn\eIIIvii{
\Det (1 \pm {\bf W} )
= \sum_{N=0}^\infty \frac{(\pm 1)^N}{N!}
\sum_{\ep_1 \cdots \ep_N = \pm }
\int_0^\infty du_1 \cdots du_N
{}~ \det \{ W_{\ep_i , \ep_j} (u_i , u_j ) \} , }
where
$\det \{ W_{\ep_i , \ep_j } (u_i , u_j ) \} $ is
an ordinary determinant of the finite $N\times N$ matrix
with $W_{\ep_i , \ep_j} (u_i , u_j ) $ as entries:
\eqn\eIIxix{
\det \left\{ W_{\ep_i , \ep_j } (u_i , u_j ) \right\}
=
\left\vert\matrix{
W_{\ep_1 \ep_1} (u_1 ,u_1 ) &W_{\ep_1 \ep_2} (u_1 ,u_2) &\cdots
&W_{\ep_1 ,\ep_N} (u_1 , u_N ) \cr
W_{\ep_2 \ep_1} (u_2 , u_1 ) &\cdot &\cdots &W_{\ep_2 \ep_N} (u_2 ,
u_N ) \cr
\cdot &\cdot &\cdots &\cdot \cr
\cdot &\cdot &\cdots &\cdot \cr
W_{\ep_N \ep_1} (u_N , u_1) &\cdot &\cdots &W_{\ep_N \ep_N} (u_N ,
u_N ) \cr}
\right\vert . }
Let $\{ \inv{u_i + v_j} \}$ for $1 \leq i,j \leq n $ denote the
$n\times n$
matrix with $1/(u_i + v_j )$ as entries.  Then using the identity
\eqn\eIIIix{
\det \{ \inv{u_i + v_j } \}
= \frac{ \prod_{i<j} (u_i - u_j )(v_i - v_j ) }{\prod_{i,j} (u_i +
v_j ) }, }
one finds that the sum \eIIIvi\ can be reproduced with the following
choice of kernel:
\eqn\eIIIx{
{\bf W} (u,v)   = \left( \matrix{
\frac{ e_+ (u) e_+ (v)}{u+v} &
\frac{ e_+ (u) e_- (v)}{u-v} \cr
\frac{ e_- (u) e_+ (v)}{u-v} &
\frac{ e_- (u) e_- (v)}{u+v} \cr } \right) , }
where
\eqn\eIIIxb{
\eqalign{
e_+ (u) &=\sqrt{\frac{f_+ (u)}{\pi} } \exp \( -\frac{m}{2} (zu +
\zbar /u ) \)
\cr
e_- (u) &= \sqrt{ \frac{f_- (u)}{\pi} } \exp \( +\frac{m}{2}
( zu + \zbar /u )   \) .  \cr}}

In principle, all of our subsequent results for $\tau_{\pm}$ can be
derived from the above formulation in terms of Fredholm determinants.
However,
the $2 \times 2$ matrix structure makes the manipulations quite
cumbersome,
and the following trick for dispensing with the $2 \times 2$
structure is quite
useful. Notice that in \eIIIvi\
the double sum over particles and holes
can be traded for an integral over $u$ running over the whole
real axis
\eqn\eIIIvin{\eqalign{
\tau_{\pm} =&
\sum_{N=0}^\infty
\frac{(\pm 1)^N}{N! }
\int_{-\infty}^\infty
\frac{du_1}{2\pi |u_1|}
\cdots \frac{du_N}{2\pi |u_N| }
\[ \prod_{i=1}^N f (u_i) e^{- (mzu_i +   m \zb /u_i ) } \]\cr
&~~~~~~~~~~~~~~~\times\[
\prod_{i<j} \( \frac{u_i - u_j }{u_i + u_j} \)^{2}  \]
.\cr}}
Here, we introduced
\eqn\fdef{f(u)={1\over 1+e^{-m(u+1/u)/2T}}.}
Now, one has
\eqn\eiiiii{
\tau_\pm  = {\rm Det} (1 \pm {\bf K} ), }
where ${\bf K}$ is a scalar kernel
\eqn\eIIIxpp{
{\bf K} (u,v)   =
\frac{ e (u) e (v)}{u+v}, }
with
\eqn\eIIIxb{
e (u) =\sqrt{\frac{f (u) {\rm sgn}(u) }{\pi} }
 \exp \( -\frac{m}{2} (zu + \zbar/u )
 \)  ,}
and it is implied that the Fredholm determinant in \eiiiii\ is taken
over functions in
the domain $-\infty < u < \infty$. There is a non-analyticity in the
kernel
at $u=0$, and any subtleties associated with this have to be resolved
by regarding the
above formalism simply as a convenient shorthand for the
better-defined $2 \times 2$
matrix integral operators in \eIIIvic. In practice, we will find that
one can simply
ignore the singular behavior at $u=0$: particles (or holes) with
small $u$,
have a very large energy ($\sim 1/u$) and their contribution is
exponentially
suppressed.

In Section 4 we will derive some differential equations that
are obeyed by  \eIIIvin.

\newsec{Ising model  with a boundary and a temperature.}

\def\rh{\hat{R}}

It is also interesting to include a finite temperature in the quantum
Ising
model
on the half line  $x\in [-\infty,0]$ with a boundary magnetic field.
This provides a good example of the convenience of  the thermal
approach.

It will be useful to think of this problem
from an Euclidean point of view, where the theory
at finite temperature is interpreted as  a theory on a cylinder of
radius $R=1/T$.
One can then consider  both possible directions as imaginary  time
direction;  we will refer to the picture where the hamiltonian
evolves
in the $x$ direction along the length of the cylinder as the L
channel,
and the crossed channel where
the evolution is along the circumference of the cylinder as the
R-channel.

At $T=0$, the boundary problem is very conveniently addressed
in the L-channel.
In this picture the boundary interactions are contained in a boundary
state $|B>$ at x=0.
Introduce the
quantity \GZ\
\eqn\zamok{\rh (\beta)=i\tanh(\beta/2) {\kappa+\cosh\beta\over
\kappa-\cosh\beta},}
which is related to the reflection matrix by $\rh
(\beta)=R({i\pi\over
2}-\beta)$.  The constant $\kappa$ is related to the boundary
magnetic field $h$ by $\kappa = 1 - h^2/2m$.
One has then \GZ\
\eqn\rtti{|B>=\exp\left[\int_{-\infty}^\infty {d\beta\over 4\pi}
 \rh (\beta)A^\dagger(-\beta)A^\dagger(\beta)\right]|0>.}

Suppose now that we include a temperature. As we have seen in section 2,
in the crossed R-channel, the effect of this temperature can be absorbed
by
defining a dressed theory, with two sorts of particles, new
form-factors,
together with an integration metric $f_\epsilon$. If one formally
carries out
the same arguments as in \GZ, one ends up with a new boundary state
that is very similar to \rtti, for the dressed theory (that is
in particular the theory involving dressed form-factors):
\eqn\bconj{
|B_T>=\exp\left[\sum_{\epsilon=\pm} \int_{-\infty}^\infty
{d\beta\over 4\pi}
\rh_\epsilon(\beta) f_\epsilon\left(i{\pi\over 2}-\beta-i\eta\right)
A^\dagger_\epsilon(-\beta)A^\dagger_\epsilon(\beta) \right]|0_T>,
}
where
\eqn\massanal{\rh_+(\beta)=\rh (\beta),
\rh_-(\beta)=\rh (\beta-i\pi),}
and the $A^\dagger_\epsilon$ operators as the same as in section 2.
In \bconj, $\eta>0$ is a shift in the imaginary rapidity
direction necessary to suppress the poles,
and ensure the right $T\to 0$ limit.

For completeness, formula \bconj\ is  proven in appendix B
using a direct approach.

As an application, let us   look at the energy
in the Ising model with boundary and temperature. Previous formula
give
\eqn\enbi{\eqalign{
<0_T|\epsilon(x)|B_T>={m\over
2}\sum_{\epsilon=\pm}\int_{-\infty}^\infty
{d\beta\over 4\pi} \hat{R}_\epsilon(\beta)
f_\epsilon\left({i\pi\over 2}-\beta-i\eta\right)\cr
e^{2mx\epsilon\cosh\beta}(e^\beta-e^{-\beta}).\cr}}

Observe that the integrals involving holes
with $\epsilon=-1$ in \enbi\
converge only when  $2|x|<\eta/T$, which implies  $T|x|<<1$. The
correlators are better
defined by  changing the contour of integration,
which is actually equivalent to a computation
in the crossed channel. For brevity,  we shall not do so here.

We can similarly consider the spin correlator. One finds

\eqn\ffoppair{\eqalign{
<0_T|\sigma|-\beta_1,\beta_1,\cdots
&,-\beta_n,\beta_n>_{\epsilon_1,...,
\epsilon_n}=\cr & (-i)^n\prod_{i=1}^n \tanh \beta_i \prod_{i<j}
\tanh\left( {\beta_i-\beta_j\over 2}\right)^{2\epsilon_i\epsilon_j}
\tanh\left( {\beta_i+\beta_j\over
2}\right)^{2\epsilon_i\epsilon_j}.\cr
}}
{}From this and \bconj\ it follows that
\eqn\tough{\eqalign{
&~~~~~~~~~~~~~~~~~<0_T|\sigma(x)|B_T>=\cr
&\sum_{n=0}^\infty {(-i)^n\over
n!}\sum_{\epsilon_i}\int_{-\infty}^\infty
\prod_i\left\{{d\beta_i\over 4\pi}
\rh_{\epsilon_i}(\beta_i)f_{\epsilon_i}\left({i\pi\over
2}-\beta_i-i\eta\right)
e^{2mx\epsilon_i\cosh\beta_i}\right\}\cr
&~~~~~~~~~~~~~~~~\times\prod_{i=1}^n \tanh \beta_i \prod_{i<j}
\tanh\left( {\beta_i-\beta_j\over 2}\right)^{2\epsilon_i\epsilon_j}
\tanh\left( {\beta_i+\beta_j\over
2}\right)^{2\epsilon_i\epsilon_j}.\cr}}
%
%
%
%
The reality of the one point
function of $\sigma$ can be checked on \tough\
by observing that
$\rh_\epsilon^*(\beta)=\rh_\epsilon
(-\beta),R_\epsilon^*(\beta)=R_\epsilon(-\beta)$.

Once again it is convenient to re-express \tough\ into
a simpler form by introducing the variable $u=e^\beta$ and
trading the sum over $\epsilon$ for an extended integral of $u$
over the whole real axis. One finds then
\eqn\final{\eqalign{
<0_T|\sigma(x)|B_T> =\sum_{n=0}^\infty
 \frac{ (-i)^n}{n!} \int_{-\infty}^\infty \prod_i
& \left\{{du_i\over 4\pi |u_i|} \rh (u_i) f (i/u_i)
e^{m x (u_i + 1/u_i )}
{u_i^2-1\over
u_i^2+1}\right\} \cr
& \times
\prod_{i<j}\left[
{u_i-u_j\over u_i+u_j}\quad {u_iu_j-1\over u_iu_j+1}\right]^2
,\cr}}
where $f (u)$ is as before \fillingu.

As was done for zero temperature in
\ref\rkon{R. Konik, A. LeClair, and G. Mussardo,  hep-th/9508099,
to appear in Int. J. Mod. Phys.},
one can express \final\ as a Fredholm determinant.
One has the identity,
\eqn\eidentit{\eqalign{
\prod_{i<j}\left[
{u_i-u_j\over u_i+u_j}\quad {u_iu_j-1\over u_iu_j+1}\right]^2
&= \prod_{i<j} \( \frac{\mu_i - \mu_j}{\mu_i + \mu_j} \)^2
\cr
&= {\rm det}  \left\{ \frac{ 2 \sqrt{\mu_i \mu_j} }{\mu_i + \mu_j }
\right\}
, \cr }}
where $\mu \equiv (u+1/u)/2$.
Thus,
\eqn\ebdet{\eqalign{
<0_T | \sigma(x) |B_T > &= {\rm Det} (1 + {\bf V})
\cr
&= \sum_{n=0}^\infty \int_{-\infty}^\infty \( \prod_i du_i \)
{\rm det} V(u_i , u_j ) , \cr }}
where ${\bf V}$ is the scalar kernel:
\eqn\ekernelv{\eqalign{
{\bf V} (u , u') &= \frac{ e(u) e(u')}{u+ 1/u + u' + 1/u' } \cr
e(u ) &=
\sqrt{ \frac{(u-1/u) (-i \hat{R} (u) ) f(i/u) }{2 \pi |u|} }
e^{mx(u+1/u)/2} . \cr }}

Using the techniques of this paper it is also possible to compute
correlation functions in boundary field theory where the finite
size is in the $x$-direction, i.e. on a infinitely long, finite
width strip.  This is described in Appendix C.

\newsec{Differential Equations for Bulk Correlators}

In this section we return to a study of the bulk two-point functions
of spin and
disorder operators at finite temperature. Recall that in Section 2 we
 had
obtained an expression for these as a Fredholm determinant of a
scalar kernel
${\bf K}$ defined in
\eIIIxpp. In this section we will use this expression to derive a set
of
non-linear, integrable, partial differential equations for the
two-point
functions. The general method we follow is that reviewed in \Korbook\
for the
case of the non-relativistic Bose gas, although we shall take a
somewhat more
algebraic point of view. The computations are somewhat involved, and
readers not
interested in the methodology can skip ahead to Section 4.5, where we
give a
self-contained presentation of the main results.

\bigskip

\subsec{The Resolvent}

\def\R{{\bf R}}
\def\H{{\bf H}}
\def\F{{\bf F}}
\def\C{{\bf C}}
\def\K{{\bf K}}

The first step in deriving differential equations for \eIIIxpp\ is to
find the
``resolvent''
for the kernel ${\bf K}$.
We will need the resolvents ${\bf R}_{\pm}$ defined by
\eqn\eIIIxi{
(1 - {\bf R}_{\pm})(1 \pm {\bf K} ) = 1. }
The resolvents ${\bf R}_{\pm}$ are also integral operators, e.g.
\eqn\eIIIxii{
 (\R \K )(u,v) = \int_{\-\infty}^\infty dw ~\R (u,w) \K (w,v) . }
The resolvent  for a kernel of the type \eIIIxpp\ was described
in \ref\rbl{D. Bernard and A. Leclair,
Nucl. Phys. B {\bf 426} (1994) 534.}.
Let $[V_\pm]$ denote the class of scalar integral operators of the
type
$e(u) e'(v)/(u\pm v)$.  As explained in\rbl,
these operators form a group with a
$Z_2$ graded multiplication law:
\eqn\eIIIxiii{
[V_{(-)} ] \times [V_{(-)} ] = [V_{(-)} ] , ~~~~~
[V_{(+)} ] \times [V_{(-)} ] = [V_{(+)} ] , ~~~~~
[V_{(+)} ] \times [V_{(+)} ] = [V_{(-)} ] . }
It turns out that $\R_\pm$ has both $[V_+], [V_-]$ pieces, so let us
write
\eqn\eIIIxvii{
\R_\pm = \H \pm \F , ~~~~~ \H \in [V_-] , ~ \F \in [V_+] , }
which turns out to be consistent with the $Z_2$ gradings.
Define functions $g_\pm (u)$ as follows:
\eqn\eIIIxxv{
( 1 - \R_\pm ) e = g_\pm .}
Above, $(\R e)(u)$ is shorthand for $\int dv \R(u,v) e (v) $.
Then the definition \eIIIxi\ and the fact that
$\K^T = \K $, $\R^T = \R$ also give us the relations
\eqn\eIIIxxva{
\eqalign{
(1 \pm \K ) g_{\pm} &= e \cr
e (1 - \R_{\pm}) &= g_{\pm} \cr
g_{\pm} ( 1 \pm \K) &= e \cr.}}
Then the resolvent can be expressed in terms of the $g$ functions:
\eqn\eIIIxxvii{\eqalign{
\H (u,v) &= \inv{2(u-v)} \( g_- (u) g_+ (v) - g_+ (u) g_- (v) \) \cr
\F (u,v) &= \inv{2(u+v)} \( g_- (u) g_+ (v) + g_+ (u) g_- (v) \)
.\cr }}

It is evident that the functions $e(u)$ in $\K$ have the 
same $z, \zbar$ dependence as for the zero temperature case
considered in \rbl, thus the sinh-Gordon differential equations
involving only $z , \zbar$ derivatives obtained there continue
to hold at finite temperature.  See also \ref\rtracy{C. Tracy
and H. Widom,  {\it 
Fredholm determinants and the mKdV/sinh-Gordon hierarchies},
solv-int/9506006.}.  
For completeness, these results will be included in the sequel.

\subsec{Intertwining Relations}

\def\dz{\partial_z}
\def\dzb{\partial_{\zbar}}
\def\d{{\bf d}}

A central property of the integral operators $\K$ and $\R_\pm$
is a set of intertwining relations between them and a family of
diagonal
operators. Essentially all of the differential equations and
constraints that we shall
derive in subsequent sections
are a simple consequence of these intertwining relations. These
relations
constitute a succinct algebraic encapsulation of the integrability of
the
$\K$ kernel.

The simplest of the intertwining relations are with the diagonal
operators $\d_n (u,v)$ ($n$ integer) which multiply the operand
by $u^n$ when $u=v$; explicitly the $\d_n$ are defined by
\eqn\essIi{\eqalign{
({\bf O}_1 \d_n {\bf O}_2)(u,v) &=
\int_{-\infty}^{\infty} dw_1 dw_2 {\bf O}_1 (u,w_1) \d_n (w_1, w_2)
{\bf O}_2 (w_2 , v) \cr
&= \int_{-\infty}^{\infty} dw {\bf O}_1 (u,w) w^n {\bf O}_2 (w,v)
\cr}}
where ${\bf O}_1$ and ${\bf O}_2$ are arbitrary operators.
Now, using the definition \eIIIx\ , we can easily
obtain our first set of intertwining relations:
\eqn\essIii{\eqalign{
(\d_1 \K + \K \d_1)(u,v) &= e (u) e (v) \cr
(\d_{-1} \K + \K \d_{-1})(u,v) &= \frac{e (u)}{u} \frac{e (v)}{v} \cr
}}
Notice that the right-hand-sides are all projection operators {\it
i.e.\/}
they are all products of functions of $u$ and $v$.
The intertwining of $\d_n$ for general $n$ with the $\K_\pm$ now
follows by
repeated application of the above and the identity $\d_{n+m} = \d_{n}
\d_{m}$.
One can also check that the operator $\d_1 \d_{-1} = 1$ commutes with
the $\K_{\pm}$.

In a similar manner we can obtain the intertwining of $\R_\pm$ with
the $\d_n$ by using \eIIIxvii\ and \eIIIxxvii:
\eqn\essIiib{\eqalign{
(\d_1 \R_{\pm} - \R_{\mp} \d_1)(u,v) &= \pm g_{\mp} (u) g_{\pm} (v)
\cr
(\d_{-1} \R_{\pm} - \R_{\mp} \d_{-1})(u,v) &= \pm \frac{g_{\pm}
(u)}{u}
\frac{g_{\mp} (v)}{v}
\cr }}
Again, the right-hand-sides are projectors.

We now introduce our second set of diagonal operators: $\d_z$ and
$\d_{\zbar}$,
which take derivatives with respect to $z$ and $\zbar$ respectively.
They are defined by
\eqn\essIiii{\eqalign{
({\bf O}_1 \d_z {\bf O}_2)(u,v) &=
\int_{-\infty}^{\infty} dw {\bf O}_1 (u,w) \partial_z {\bf O}_2 (w,v)
\cr
({\bf O}_1 \d_{\zbar} {\bf O}_2)(u,v) &=
\int_{-\infty}^{\infty} dw {\bf O}_1 (u,w) \partial_{\zbar} {\bf O}_2
(w,v) \cr
}}
Notice that the derivatives act on everything to their right {\it
i.e.\/}
on the ${\bf O}_2$ and on all operands the left-hand-side may act on.
Again, using the defining relations \eIIIx, \eIIIxb, we can
easily obtain the intertwining of the $\d_z$, $\d_{\zbar}$ with
$\K$:
\eqn\essIiv{\eqalign{
(\d_z \K - \K \d_z)(u,v) &= - \frac{m}{2} e (u) e (v) \cr
(\d_{\zbar} \K - \K \d_{\zbar})(u,v) &= - \frac{m}{2}\frac{e (u)}{u}
\frac{e (v)}{v}
\cr }}
As usual, the right-hand-sides are projectors.
The intertwining of the $\R_\pm$ with the $\d_z$, $\d_{\zbar}$ can be
deduced
from \essIiv\ and \eIIIxi,
but as we shall not need it in this paper, we will refrain from
displaying it.

Finally, we will need the intertwining of $\K_\pm$ with an operator
involving
temperature derivatives. We shall follow the idea of \Korbook, and
choose an
operator which commutes with the thermal factor \fillingu. The
appropriate operator,
$\d_T$ is defined by
\eqn\essIv{
({\bf O}_1 \d_T {\bf O}_2)(u,v) =
\int_{-\infty}^{\infty} dw {\bf O}_1 (u,w) \left(
\left(w - \frac{1}{w} \right) T \frac{\partial}{\partial T} +
\left(w + \frac{1}{w} \right) w \frac{\partial}{\partial w} \right)
 {\bf O}_2 (w,v).}
Again all derivatives act on all operators to their right. This
particular
combination of derivatives was chosen because
\eqn\essIvi{
\left(
\left(u - \frac{1}{u} \right) T \frac{\partial}{\partial T} +
\left(u + \frac{1}{u} \right) u \frac{\partial}{\partial u} \right)
\Omega \left(
(u + 1/u)/T \right) = 0}
where $\Omega(y)$ is an arbitrary function of the combination
$y=(u+1/u)/T$
alone. This is precisely the combination in which $T$ occurs in
$\K_\pm$,
and \essIvi\ raises the possibility that the intertwining of $\d_T$
with $\K_\pm$ may be simple. It turns out that the result is somewhat
more
complicated than those obtained earlier, but is nevertheless a sum of
a small
number of projection operators.
A lengthy, but ultimately straightforward, computation gave us
\eqn\essIvii{\eqalign{
(\d_T \K + \K \d_T)(u,v) &= e (u) T \frac{\partial}{\partial
T} e (v) - \frac{ e (u)}{u} T \frac{\partial}{\partial
T} \frac{e (v)}{v} - e (u) e (v) \cr
&~~~~~~~~~~~~~~~- \frac{m}{2} (u-v) \left( z  + \frac{\zbar}{u^2 v^2}
\right) e
(u) e (v)  \cr
}}
All derivatives, including the ones appearing on the
right-hand-sides, act on
all operators to their right. This equality holds in the sense that
if both sides are multiplied by
an arbitrary function of $v,T,z,\bar{z}$, and $v$ is then integrated
over, the results are equal up to a surface term which is presumed to
vanish.
Notice especially the location of the $T$ derivatives on the
right-hand-side
-- this leads to the most compact form of the above equation, and is
also the
most convenient in subsequent manipulations; when the above equation
acts
on the $g$ function, this form immediately gives $T$ derivatives of
the
``potentials'' which we will introduce in the following subsection.

The identities \essIii, \essIiib, \essIiv, and \essIvii\ are the main
results of this section. Our results in the remainder of the paper
for the relations
obeyed by
$\tau_{\pm}$ (except for those in Appendix D) follow by simple,
repeated applications of
these intertwining relations.

\subsec{ Definition of the potentials}

We need to introduce some additional formalism and notation before
turning
to the study of the differential equations.

We consider here the so-called ``potentials'', which are the set of
possible
scalar overlaps among the $e$ and $g$ functions introduced in Section
4.1.
These occur, for instance, when we consider derivatives of the
$\tau_\pm$
correlators. Thus from \eiiiii\
\eqn\essIIi{\eqalign{
\dz \ln \tau_+ &= \dz \ln \Det( 1 + \K ) \cr
&=  \Tr \left( ( 1 + \K )^{-1} \dz \K \right) \cr
&= - \frac{m}{2} \Tr \left( ( e (1 - \R_+) e \right) \cr
&= - \frac{m}{2} \Tr \left( e g_+ \right). \cr }}
In the last step we used the defining relation \eIIIxxv.
The expression $\Tr \left( e g_+ \right)$ is one of the potentials;
notice that the variable $u$ has been integrated over, and the
potential
is an implicit function of $z$, $\zbar$, and $T$ alone.
We will also need potentials defined by inserting various powers of
the
variable
$u$ in the scalar product:
\eqn\essIIii{
\Tr \left( e  \d_n g_+ \right) = \int_{-\infty}^{\infty} du u^n
e (u)  g_+ (u).}
To make things compact, we introduce some notation here for the
potentials that turn up in our analysis.
Our main results will be expressed in terms of the following
potentials
\eqn\essIIvii{\eqalign{
b_{\pm} &= \Tr \left( e  g_{\pm} \right) \cr
c_{\pm} &= \Tr \left( e \d_{-1} g_{\pm} \right) \cr
d_{\pm} &= (1\pm c_{\mp}) \Tr \left( e \d_{-2} g_{\pm} \right) \cr
}}
The reason for the additional factors in the definitions of the
$d_{\pm}$
potentials will become clear shortly. Not all of these potentials are
independent; the intertwining relations
\essIii\ and \essIiib\ leads to a constraint between them:
\eqn\essIIv{\eqalign{
c_+ &= \Tr \left(e \d_{-1} g_+ \right) =
 \Tr \left( e  \d_{-1} (1 - \R_{+} ) e \right) ~~~~~{\rm from
\eIIIxxv\  } \cr
&~~~~~~~~~= \Tr \left( e  (1 - \R_{-} ) \d_{-1} e \right) -
c_+ c_-
{}~~{\rm from \essIiib\  }\cr
&~~~~~~~~~= \Tr \left( g_- \d_{-1} e \right) - c_+ c_-
{}~~{\rm from \eIIIxxva\  }, \cr
}}
which using \essIIii\ becomes finally
\eqn\essIIviia{c_- - c_+ = c_+ c_-.}
We solve this constraint by henceforth parameterizing
\eqn\essIIviiaa{c_+ = 1 - e^{\phi}~~~,~~~c_- = e^{-\phi} - 1.}

Let us also generalize \essIIi, and tabulate all the derivatives of
the $\tau_{\pm}$:
\eqn\essIIviii{\eqalign{
\dz \ln \tau_{\pm} &= \mp \frac{m}{2} b_{\pm} \cr
\dzb \ln \tau_{\pm} &= \mp \frac{m}{2} d_{\mp} \cr
}}
Our choices in the definitions of the $d_{\pm}$ earlier were made so
that the
above equations came out in this symmetrical form.

In our subsequent analysis, some additional potentials also arise at
intermediate
stages. To handle these, it is convenient to introduce here some
additional notation
\eqn\essIIix{\eqalign{
a_1 &= \Tr \left( e \d_{1} g_{+} \right) + \Tr \left( e \d_{1} g_{-}
\right) \cr
a_2 &= e^{-\phi} \Tr \left( e \d_{-3} g_{+} \right) + e^{\phi} \Tr
\left( e \d_{-3}
g_{-}
\right). }}
Both $a_1$, $a_2$ will drop out of our final results.

\subsec{Differential Equations}

We will first consider the differential equations which involve only
derivatives with respect to $z$ and $\zbar$. These will turn out to
be closely
related to those derived earlier by Perk {\it et. al.\/} by a very
different
computation on a lattice model. Then we will turn to a new set of
equations
involving temperature derivatives.

\bigskip
\noindent
{\it 4.4.1 Spacetime derivatives}

We begin by turning the intertwining relations \essIiv\ into
differential
equations for the $f$ functions. Begin with the equation $(1 + \K )
g_+ = e$ (\eIIIxxva) and act on both sides
with the operator $(1 - \R_+ ) \d_z$. Simplifying this and related
equations by
repeated use of all the intertwining relations gives us
\eqn\essIIIi{\eqalign{
\d_z \( \matrix{ g_+ \cr g_- \cr} \) (u) &=
- \frac{m}{2} \( \matrix{ -b_+ - b_- & u \cr u & b_+ + b_- \cr} \)
\( \matrix{ g_+ (u) \cr g_- (u) \cr} \) \cr
\d_{\zbar} \( \matrix{ g_+ \cr g_- \cr} \) (u) &=
- \frac{m}{2} \( \matrix{ 0 & e^{2\phi}/u \cr e^{-2 \phi}/u & 0 \cr}
\)
\( \matrix{ g_+ (u) \cr g_- (u) \cr} \). \cr
}}
These are a set of equations in the Lax form, and integrability
condition
$\d_z \d_{\zbar} = \d_{\zbar} \d_z$ leads to differential equations
among the
potentials alone. However, a richer set of equations is obtained by
acting
on the definitions of the potentials \essIIvii\ by $\d_z$ and
$\d_{\zbar}$,
using \essIIIi, and the derivatives of the $e$ functions which follow
from
their definition in \eIIIxb. This gives us
\eqn\essIIIvi{\eqalign{
\partial_z \phi &= \frac{m}{2} ( b_+ + b_- )  \cr
\partial_{\bar{z}} \phi &= \frac{m}{2} ( d_+ + d_- )
\cr
\partial_{\bar{z}} b_{\pm} &= m e^{\pm \phi} \sinh \phi  \cr
\partial_{z} d_{\pm} &= m e^{\mp \phi} \sinh \phi . \cr
}}
With these equations in hand, we can now express \essIIIi\ in a more
symmetrical form.
First define the functions $h_{\pm} (u)$ by
\eqn\essIIIvii{
h_{\pm} (u) = e^{\mp \phi/2} g_{\pm} (u).}
Note that $e(u)$, $g_{\pm} (u)$, and $h_{\pm} (u)$ are the only
functions which depend
upon the spectral parameter $u$; all other functions are independent
of $u$.
Now it is easy to show from \essIIIi\ and \essIIIvi\ that
\eqn\essIIIviii{\eqalign{
\d_z \( \matrix{ h_+ \cr h_- \cr} \) (u) &=
- \frac{1}{2} \( \matrix{ -\partial_z \phi & m ue^{-\phi} \cr u
e^{\phi} & \partial_z
\phi \cr} \)
\( \matrix{ h_+ (u) \cr h_- (u) \cr} \) \cr
\d_{\zbar} \( \matrix{ h_+ \cr h_- \cr} \) (u) &=
- \frac{1}{2} \( \matrix{ \partial_{\zbar} \phi & m e^{\phi}/u \cr m
e^{- \phi}/u &
- \partial_{\zbar} \phi \cr}\)
\( \matrix{ h_+ (u) \cr h_- (u) \cr} \). \cr
}}

{}From \essIIviii\ and \essIIIvi\ , it can be shown that
the second derivatives of the $\tau$ functions are characterized by
the
compact equations:
\eqn\etaub{\eqalign{
\partial_z \partial_\zbar \ln \tau_\pm &= \mp \frac{m^2}{2} e^{\pm
\phi}
\sinh \phi \cr
\partial_z \partial_\zbar \phi &= \frac{m^2}{2} \sinh 2\phi . \cr }}
The second of these equations is the compatibility condition $\d_z
\d_{\zbar}
= \d_{\zbar} \d_{z}$ of \essIIIviii.

Another simple consequence of \essIIIvi\ follows from combining it
with the definitions
\essIIviii; one notices easily that
\eqn\essni{
\partial_z \left( \ln (\tau_{-}/\tau_{+}) - \phi \right) = 0~~~~~~
\partial_{\zbar} \left( \ln (\tau_{-}/\tau_{+}) - \phi \right) = 0,}
which implies that $\phi = \ln (\tau_{-}/\tau_{+}) + $ a
dimensionless constant
dependent only on the ratio $T/m$. In Appendix D we show that the
integration constant is
in fact also independent of $T/m$, and then show by an explicit
computation at $T=0$
that its value is zero. Therefore we have
\eqn\essnii{
e^{\phi} = \frac{\tau_{-}}{\tau_{+}}.}

It is also useful to consider differential equations satisfied by the
$a_{1,2}$ potentials defined in \essIIix; as we will now show, these
equations lead to
additional differential equations involving only the $b_{\pm}$ and
$d_{\pm}$.
Using the same methods as those used for \essIIIvi\ we can obtain
\eqn\essIIIix{\eqalign{
\partial_z b_{\pm} &= - \frac{m}{2} \left( a_1 \mp b_{\pm} (b_+ + b_-
) \right) \cr
\partial_{\zbar} d_{\pm} &= - \frac{m}{2} \left( a_2 \pm d_{\pm} (d_+
+ d_- ) \right)
\cr
\partial_{\bar{z}} a_1 &= - m  ( b_+ e^{-\phi} + b_- e^{\phi} ) \cosh
\phi \cr
\partial_{z} a_2 &= - m  ( d_+ e^{\phi} + d_- e^{-\phi} ) \cosh \phi
\cr
}}
The last two equations above are in fact not new; we can solve the
first two for
$a_{1,2}$ and then they follow from applications of \essIIIvi.
We can also eliminate $a_{1,2}$ between the first two equations of
\essIIIix\
and obtain
\eqn\essIIIx{\eqalign{
\partial_z b_+ - \partial_z b_- &= \frac{m}{2} \left( b_+ + b_-
\right)^2 \cr
\partial_{\zbar} d_+ - \partial_{\zbar} d_- &= - \frac{m}{2} \left(
d_+ + d_- \right)^2
\cr
}}
We will not use the $a_{1,2}$ any more in this paper, although we
will implicitly use
the first two equations in \essIIIix\ in the derivation of some
equations in the next
section.

\bigskip
\noindent
{\it 4.4.2 Temperature derivatives}

We will now follow the same procedure as in 4.4.1, but with the
$\d_{T}$
operator replacing the $\d_z$ and $\d_{\zbar}$ operators.

First, it is useful to notice that
\eqn\essIVi{
(\d_{T} e )(u) = -\frac{m}{2}\left( u + \frac{1}{u} \right) \left( z
u -
\frac{\bar{z}}{u}
\right) e (u) .}
We can now get the $\d_{T}$ derivatives of the $f$ functions by
acting on
the $(1 \pm \K) g_{\pm} = e$ equations by $(1-\R_{\mp}) \d_{T}$ on
the
left; this gives
\eqn\essIViia{
\d_{T} \( \matrix{ g_+ \cr g_- \cr} \) (u) =
 \( \matrix{ A & B_+ \cr B_- & -A} \)
\( \matrix{ g_+ (u) \cr g_- (u) \cr} \)}
where
\eqn\essIVii{\eqalign{
A &=  -\frac{T}{u} \partial_T \phi
 + \frac{muz}{2} ( b_+ + b_- )
+ \frac{m\bar{z}}{2 u} (d_- + d_+ ) \cr
B_+ &=  \( 1 + z \partial_z -
T \partial_T \)
b_+  -
\frac{mz}{2}
\left( u^2 + 1 \right) + \frac{m
\bar{z}}{2 u^2}
\left(  u^2 + e^{2\phi}
\right)  \cr
B_- &=  -\( 1 + z \partial_z - T \partial_T \)
b_- - \frac{mz}{2} \left( u^2 + 1\right) + \frac{m
\bar{z}}{2 u^2}
\left(  u^2 + e^{-2\phi}
\right) ; \cr
}}
we have used \essIIIix\ to eliminate the $a_1$ potential which arises
in the above
derivation. We will write this equation in a symmetrical form
momentarily.

First, we derive equations relating the $T$ derivatives of the
potentials;
we will do this by evaluating
\eqn\essIViii{
\Tr \left( \d_{-1} \d_{T} ( e g_{\pm} ) \right)}
in two different ways. For the first we use $\d_{T} (e g_{\pm}) =
g_{\pm}
\d_{T} e + e \d_{T} g_{\pm}$ and evaluate the right hand side using
\essIVi\ and \essIViia; alternatively, we can write, using the
definition
\essIv,
\eqn\essIViv{
\Tr \left( \d_{-1} \d_{T} ( e g_{\pm} ) \right) =
\left(  T \partial_T -1 \right)
\Tr \left( e ( 1 - \d_{-2}) g_{\pm}  \right),}
where we have integrated by parts over the $u$ integral. Comparing
the results of
these two evaluations, we obtain
\eqn\essIVv{
e^{\mp \phi} \( 1 + z \partial_z - T \partial_T \) b_{\pm} -
e^{\pm \phi} \( 1 + \zbar \partial_{\zbar}- T
\partial_T \) d_{\pm} = m(\zbar - z) \sinh \phi}
Again, we have used \essIIIix\ to eliminate the $a_{1,2}$ potentials
at
intermediate stages.

Now we can combine \essIIIvii and \essIVv\ to put \essIViia\ into a
more symmetrical
form:
\eqn\essIVvi{
\d_{T} \( \matrix{ h_+ \cr h_- \cr} \) (u) =
 \( \matrix{ \tilde{A} & \tilde{B}_+ \cr \tilde{B}_- & -\tilde{A}} \)
\( \matrix{ h_+ (u) \cr h_- (u) \cr} \)}
where
\eqn\essIVvii{\eqalign{
\tilde{A} &=  -\frac{1}{2} \( u + \frac{1}{u} \) T \partial_T \phi
 + \frac{muz}{2} ( b_+ + b_- )
+ \frac{m\bar{z}}{2 u} (d_- + d_+ ) \cr
\tilde{B}_{\pm} &= \pm \frac{e^{\mp \phi}}{2} \( 1 + z \partial_z -
T \partial_T \)
b_{\pm}  \pm \frac{e^{\pm \phi}}{2} \( 1 + \zbar \partial_{\zbar} -
T \partial_T \)
d_{\pm} \cr
&~~~~~~~~~~~~ -
\frac{m}{2}
\left( z u^2 e^{\mp \phi} - \frac{\zbar e^{\pm \phi}}{u^2} \right) -
\frac{m}{2}
(z - \zbar ) \cosh \phi . \cr
}}
This is another set of equations in the Lax form, and like
\essIIIviii\ they behave
simply under the $z \leftrightarrow \zbar$, $u \leftrightarrow 1/u$,
$b_{\pm}
\leftrightarrow d_{\pm}$  transformations. We can examine  the
consequences of the compatibility conditions
$\d_{T} \d_z = \d_z \d_{T}$ and $\d_{T} \d_{\zbar} = \d_{\zbar}
\d_{T}$ on \essIVvii\ and  \essIIIviii; this generates equations for
the potentials
which turn to be already implied by the equations in Section 4.4.1
and \essIVv.

\subsec{Recapitulation}

Our derivation of the differential equations has been rather
circuitous, so it seems
useful here to present a self-contained review of the final results.
The discussion
here is also designed to be accessible to readers who have skipped
Sections 4.1-4.4.

The main quantities of interest are the correlators $\tau_{\pm}$
defined in \eIIIv.
We now introduce the variables $b_{\pm}$, $d_{\pm}$ and $\phi$ by
\eqn\erci{
\eqalign{
b_{\pm} & = \mp \frac{2}{m} \partial_z \ln \tau_{\pm} \cr
d_{\pm} & = \pm \frac{2}{m} \partial_{\zbar} \ln \tau_{\mp} \cr
\phi &= \ln ( \tau_- / \tau_+ ). \cr
}}
These variables were introduced earlier as ``potentials'', but here
we shall simply
consider \erci\ as their defining relations.

A complete, mutually-independent, set of non-linear partial
differential equations
obeyed by these quantities is
\eqna\ercii
$$\eqalignno{
\partial_z \partial_\zbar \ln \tau_\pm &= \mp \frac{m^2}{2} e^{\pm
\phi}
\sinh \phi &\ercii a\cr
\partial_z b_+ - \partial_z b_- &= \frac{m}{2} \left( b_+ + b_-
\right)^2 &\ercii b \cr
\partial_{\zbar} d_+ - \partial_{\zbar} d_- &= - \frac{m}{2} \left(
d_+ + d_-
\right)^2
&\ercii c\cr
e^{\mp \phi} \( 1 + z \partial_z - T \partial_T \) b_{\pm} & -
e^{\pm \phi} \( 1 + \zbar \partial_{\zbar}- T
\partial_T \) d_{\pm}  = m(\zbar - z) \sinh \phi .
&\ercii d\cr}
$$
The equations \ercii{b}, \ercii{c}, and the difference of the two
equations in
\ercii{a}\  precisely exhaust the set of equations obtained earlier
in
\mccoy,\perk. Note that these equations do not involve $T$
derivatives, and we
have one more equation of this type (one of the two in \ercii{a})
than those
obtained earlier. The equations in \ercii{d}\ are new.

The equations in \ercii{a}\ and \ercii{d}\
can also be written in the Lax form\ref\caveat{
The equations in the Lax form actually contain slightly less
information than that in
\ercii{a}\ and \ercii{d}. Let us write $L_{\pm} = \partial_z
\partial_\zbar \ln \tau_\pm \pm
(m^2 /2) e^{\pm \phi}
\sinh \phi$. Then the Lax form actually only implies $L_+ - L_- = 0$,
$(1 + z \partial_z - T \partial_T ) L_+ = 0$, and $(1 + \zbar
\partial_{\zbar} - T
\partial_T ) L_- = 0$, which is slightly weaker than \ercii{a}\ which
implies
$L_{\pm} = 0 $. The equations \ercii{d}\ however emerge completely
from the Lax form.}
{\it i.e.\/} as the integrability conditions of linear equations
obeyed by  auxilliary
functions dependent upon an additional spectral parameter. We need 2
auxilliary functions
$h_{\pm} (u)$, where
$u$ is the spectral parameter. The linear equations satisfied by
$h_{\pm} (u)$ are
\essIIIviii\ and \essIVvi, where we note that on the left-hand-sides
of these
equations one can replace $\d_z \rightarrow \partial_z$,
$\d_{\zbar} \rightarrow \partial_{\zbar}$ and $\d_T \rightarrow
(u-1/u) T \partial_T + (u+1/u) u \partial_u $.

\subsec{Massless Limit}

In this section we will examine the behavior of the non-linear
partial differential
equations in the limit of small $m/T$. We recall
 that all correlators are smooth functions of $m$ at $m=0$
for finite $T$ \ref\AlyoZam{A. Zamolodchikov, Nucl. Phys. B {\bf 348}
(1991) 619},\subir, so
the
$\tau_{\pm}$ obey the following series expansions
\eqn\emli{
\eqalign{
\( \frac{m}{T} \)^{1/4} \tau_+ & = \sum_{n=0}^{\infty}  \(
\frac{m}{T}
\)^{2n} \tau_{2n} (zT , \zbar T)  \cr
\( \frac{m}{T} \)^{1/4} \tau_- & = \sum_{n=0}^{\infty}  \(
\frac{m}{T}
\)^{2n+1} \tau_{2n+1} (zT , \zbar T).\cr}}
The prefactor of $(m/T)^{1/4}$ follows from the overall normalization
condition
for $\tau_{\pm}$ implicit in \eIIIvi, and the behavior of the
correlators at
$m=0$. The coefficients $\tau_n$ in \emli\ are dimensionless
functions
only of
$z T$ and $\zbar T$. Indeed, the $\tau_n$ are clearly spacetime
integrals over
correlators in the massless conformal field theory of two $\sigma$ or
$\mu$ operators
and $n$ thermal (energy) operators.
These correlators possess a holomorphic/anti-holomorphic
factorization property, but
it is not expected that the result of a spacetime integral over them
in a cylinder
geometry will be in a factorized form. Let us stress here that
conformal
perturbation theory  expansions
such as
 \emli\ as possible only because $T\neq 0$, so there are no infrared
divergences
\IS.

An interesting question is the convergence of expansions
such as \emli. Of course, the absence of any phase transitions at
finite temperature in
one-dimensional physical systems with short-range interactions
implies that there
can be no singularity at any finite  real value of $m/T$.
  The only singularities  on the real axis consist
of powers of $\exp(-m/T)$ as $m/T \rightarrow \infty$ (this is clear
from the
results of \subir). However, phase
transitions are possible in one-dimensional systems
with unphysical  complex  couplings. In the present case,
the largest eigenvalue in the spin even sector
crosses the largest eigenvalue in the spin odd sector when $m=i\pi
T$,
and presumably this determines the radius of convergence
of \emli.

We can now insert \emli\ into the definitions \erci\ and the
differential
equations \ercii. All the equations organize themselves neatly in
series of
integer powers of
$m$, and demanding that they are obeyed at each order leads to a
hierarchy
of  differential equations obeyed by the $\tau_n$.
The zeroth order equations in this hierarchy involve only $\tau_0$
and they are
\eqn\emlii{
\eqalign{
\partial_z \partial_{\zbar} \ln \tau_0 & = 0 \cr
( 1 + z \partial_z - T \partial_T ) \partial_z \ln \tau_0 & = 0 \cr
( 1 + \zbar \partial_{\zbar} - T \partial_T ) \partial_{\zbar} \ln
\tau_0 & = 0. \cr
}}
As $\tau_0$ is proportional to the spin correlators of the massless,
conformal,
theory, it is expected to factorize into holomorphic/anti-holomorphic
parts:
\eqn\emliii{
\tau_0 (w , \bar{w} ) = \Psi_0 (w) \Psi_0 ( \bar{w}),}
where we have introduced $w \equiv z T$, $\bar{w} = \zbar T$.
Substituting \emliii\ for $\tau_0$, we find that \emlii\ are obeyed
for {\it any\/}
function $\Psi_0 (w)$.
The form of $\Psi_0$ follows from a, now standard, argument based on
conformal
invariance \ref\cardy{J.L. Cardy, J. Phys. A {\bf 17}, L385 (1984)},
and its
overall scale was determined in \subir:
\eqn\emliv{
\Psi_0 (w) = \frac{2^{5/12} \pi^{1/8} e^{1/8} A_G^{-3/2} }{( \sin ( 2
\pi w )
)^{1/8}},}
where $A_G$ is Glaisher's constant ($\ln A_G = 1/12 - \zeta' (-1)$).
Notice that, apart from the scale, $\Psi_0$ is determined by the
requirement
that its only singularities be $\Psi_0 ( w \rightarrow 0 ) \sim
w^{-1/8}$
and its periodic images under $w \rightarrow w + 1/2$, and that
it have no zeros.

\newsec{Conclusion}

In conclusion, we would like to stress that the differential
equations involving only spatial coordinates (the
equations \ercii{a\hbox{--}c}) would hold for any filling function
$f$ in
\eIIIvi.
This is because these equations follow
only from the monodromy relations \essIii,\essIiv, that are
insensitive
to $f$. The last
equation in \ercii\ would hold for any filling function $f$ that
depends only on $(u+1/u)/T$ and satisfies $f(\infty)=1,f(-\infty)=0$.
This
is because the differential equation follows from \essIvii\ that in
turns
follows from \essIvi, together with \essnii\ that depends on the
$T=0$ behavior
of $f$. This does not
mean that the differential equations are too general to be useful:
they
can be quite constraining after imposing some limiting behavior
of the correlators (like \emliv) determined by other means (similar
features are known for the Bose gas \Korbook).

While we have not derived the differential equations for
the boundary problem, we can straightforwardly obtain some of them.
Indeed,  the kernel \ekernelv\  differs from the  zero temperature
one in
\rkon\ by the filling function  $f$ only.  Thus, as for the bulk
two point function, the differential equations
derived in \Bar\rkon\ involving
 derivatives with respect to $\kappa$
and $x$ still apply for the non-zero temperature correlator.

The presence of a non-zero temperature also introduces a number of
new physical
phenomena, not found at $T=0$, which are of considerable
theoretical interest. Foremost among these is the appearance of
irreversible,
thermal relaxational behavior in the long-time spin correlations (for
field-theorists, it is perhaps useful to note here that we are
referring to
correlations in {\it real\/} time; this relaxational behavior has no
simple
characterization in Euclidean time where correlations are periodic
with period
$1/T$). While this relaxational behavior is expected to occur at all
non-zero
$T$, there are some important distinctions in its physical nature
between
high $T$ ($T \gg m$) and low $T$ ($T \ll m$), where $m$ is mass gap
above the
ground state of the continuum Ising model.
\vskip 1pt
\noindent
({\it i\/}) $T \gg m$:
The presence of thermal relaxational behavior has been established in
this
limit, and a simple argument based on conformal invariance also
allowed exact
computation of a relaxation rate constant
\ref\statphys{S. Sachdev in {\it Proceedings of the 19th IUPAP
International Conference on Statistical Physics\/},
Xiamen, China, July 31 - August
4 1995,  ed. B.-L. Hao, (World Scientific, Singapore 1996)}. In this
high $T$
regime, the
fluctuations are dominated by states with energy
$\hbar \omega \sim k_B T$, and the spin dynamics are therefore not
amenable to
a description by some effective classical model of dissipative
dynamics
\csy.
\vskip 1pt
\noindent
({\it ii\/}) $T \ll m$:
The physical situation is quite different at low $T$ just above the
ordered
state (we will not comment here on the behavior above the quantum
paramagnetic
state). Now the most important
excitations have an energy
$\hbar
\omega \ll k_B T$ \subir, and their de-Broglie wavelength is
therefore much
smaller than their mean spacing. This reasoning led to the conjecture
\subir\
that the long-time relaxational spin dynamics is described by an
effective 
{\it classical} model.

The study in this paper is a step towards understanding the finite
temperature dynamics of the continuum quantum Ising model, as it
crosses over
from the high $T$ limit (where the dynamic correlations are known in
closed form)
to its conjectured classical relaxational
behavior at low $T$. In all our
computations,
time evolution is specified by the usual unitary Heisenberg operator
$e^{i H t/\hbar}$, where $H$ is the Hamiltonian on the quantum Ising
model,
and there is no coupling to a heat bath. The temperature appears
solely in
specifying an initial density matrix $e^{-H/T}$, which is also
stationary under
the time evolution.  
We have taken several significant steps towards obtaining dynamic
correlation
functions under the unitary Heisenberg time evolution, and
its complete understanding 
now appears within reach.

\bigskip
\noindent {\bf Acknowledgments}: A.L. was supported by the NSF (in
part through the National Investigator program). 
F.L. and H.S.  were supported by the
 Packard Foundation, the NSF (through the National Investigator
Program), and the
DOE. F.L. was also partially supported
by a Canadian NSERC postdoctoral Fellowship. S.S. was supported by
NSF Grant
DMR 96-23181.

\vfill
\eject

\appendix{A}{Direct Approach}

\def\t{\beta}
\def\th{\beta}
\def\Phi{\CO}

In this appendix, we would like to
give a different derivation of the result \generes.

To start, we
consider the one point function of some operator \CO:
\eqn\IIi{
\<\CO\>_T = \inv{Z}  \sum_\psi ~ e^{- E_\psi /T}  {\<\psi| \CO
|\psi \>\over \<\psi|\psi\>}  .}

In a multiparticle basis for the states $| \psi >$, one has the
following
resolution of the identity when $L=\infty$:
\eqn\eIIiii{
1 = \sum_{n=0}^\infty \inv{n!} \int_{-\infty}^\infty { d\th_1 \cdots
d\th_n\over
(2\pi)^n} ~
      | \th_1 , \ldots, \th_n \> \< \th_n , \ldots , \th_1 | . }
Thus,
\eqn\eIIiv{
\<\CO\>_T = \inv{Z} \sum_{n=0}^\infty \inv{n!}
 \int d\th_1 \cdots d\th_n ~ \( \prod_{i=1}^n e^{- e(\th_i)/T } \)
       \< \th_n , \ldots , \th_1 | \CO | \th_1 , \ldots , \th_n \>  .
}

The matrix elements of $\CO$ are the so called  form-factors.  As we
will explain below, one finds
that the sum \eIIiv\ has two kinds of singularities that must be
dealt with before one can obtain a meaningful expression.
  The first kind of singularity is a result of
working with multiparticle states that are defined in infinite
volume,
i.e. for $L=\infty$.
These infinite volume singularities manifest themselves as
$\delta (0)$, i.e. $\delta (\th - \th' )$ with $\th = \th'$.  As
expected
on general grounds, these infinite volume singularities are removed
upon dividing by $Z$.  The second kind of singularity arises from
poles in the form factors.  These `pole singularities' can be removed
by a renormalization of the one-point functions.

\def\dag{\dagger}

One first needs some general properties of the form factors.
We specialize to a theory with S-matrix equal to $-1$.
In terms of the operators defined in \fadzamo,
the multiparticle states have the following representation:
\eqn\eIIv{\eqalign{
|\th_1 , \ldots , \th_n \>  &= A^\dag (\t_1 ) \cdots A^\dag (\t_n )
|0\> \cr
\<\th_n , \ldots , \th_1 | &= \< 0| A (\t_n ) \cdots A^\dag (\t_1 )
. \cr}}
Using the algebra of the operators $A , A^\dag $, one can evaluate
the inner products $\<\t_n,...| \t_1 ....\>$.

Let $\CA , \CB$ denote some ordered sets of rapidities,
$\CA = \{ \t_n , ..., \t_1 \} $, $\CB = \{ \t'_1 , ..., \t'_m \} $,
and consider the form factor $\<\CA | \Phi (x) |\CB\> $.  If
$\CA$ and $\CB$ have no overlap, i.e. $\<\CA | \CB \> = 0$, then the
form factor obeys the crossing relation (see also next appendix):
\eqn\eIIvii{
\<\CA | \CO (x) |\CB\> = \<0| \CO (x) |\CB , \CA - i\pi \> , }
where $\CA - i \pi$ denotes all rapidities shifted by $- i\pi$.
If $\<\CA| \CB \> \neq 0$, the form factor is expressed as a sum over
all ways of breaking up $\CA, \CB$ into two sets \Smir:
\eqn\eIIviii{
\<\CA | \CO (x) |\CB\>  = \sum_{\CA = \CA_1 \cup \CA_2;
\CB = \CB_1 \cup \CB_2 } ~
\< \CA^+_1| \CO (x) |\CB_1\>
\<\CA_2 | \CB_2 \> ~ S_{\CA, \CA_1} S_{\CB , \CB_1} . }
In this formula, $\CA^+ $ denotes rapidities shifted by
a infinitesimally small imaginary part $i\eta$, so that the
crossing relation is valid:
\eqn\eIIix{
\<\CA^+ | \CO | \CB \> = \< \CA + i\eta | \CO
| \CB \> =   \<0| \CO |\CB, \CA - i\pi + i\eta  \> . }
The S-matrix factors are defined as the product of S-matrix elements
required to bring $|\CA\>$ into the order $|\CA_2, \CA_1 \>$, and
similarly
for $\CB$.  Namely,
\eqn\eIIx{
\<\CA | = S_{\CA, \CA_1 } \< \CA_2, \CA_1 | , ~~~~~~~
|\CB \> = S_{\CB, \CB_1 } | \CB_1, \CB_2 \> . }
We will refer to the single term  in \eIIviii\ with $\CA_1 = \CA,
\CB_1 = \CB$
as the connected piece and the rest as the disconnected pieces of the
form factor.

We will also need the partition function
$Z$.  In infinite volume, $Z$ is singular, and it is well known how
to
regulate it on a cylinder of finite length.  It suffices for our
purposes
however to work with an unregulated expression for $Z$, since, as we
will see, dividing by $Z$ simply cancels similar infinite volume
divergences in the sum \eIIiv.
Therefore, we take
\eqn\eIIxi{
Z =  \sum_{n=0}^\infty \inv{n!}
 \int {d\th_1 \cdots d\th_n\over (2\pi)^n} ~ \( \prod_{i=1}^n e^{-
e(\th_i) /T } \)
       \< \th_n , \ldots , \th_1 |  \th_1 , \ldots , \th_n \>  . }
Up to two particles, using the Faddeev-Zamolodchikov algebra one
finds:
\eqn\eIIxii{
Z = 1 + \int d\t e^{- e(\t )/T}  <\th|\th>
  + \inv{2} \int  d\t_1 d\t_2 e^{- (e(\t_1 ) + e(\t_2 ))/T}
  <\th_2 ,\th_1 |\th_1 , \th_2 >
  + .....}

 One can explicitly evaluate the lowest order
terms in the sum \eIIiv, using \eIIviii\ and \eIIxii.  One finds that
the effect of the disconnected pieces of the form factors is
two-fold:
some of the disconnected terms lead to $\delta(0)$ singularities as
in $Z(T)$, whereas others lead to a modification of the integration
measure $\int d\th$.  Doing the explicit computation up to 3
particles,
one finds,
\eqn\eIIxiii{\eqalign{
&~~~~~~~~~~~~~~~~\sum_\psi e^{- E_\psi /T } { \<\psi | \Phi | \psi
\>\over\<\psi|\psi\>}
= Z ( \<0|\Phi |0\>\cr
& ~~~~~+ \int d\th e^{- e(\th )/T}
\( 1 -  e^{- e(\th)/T} + e^{-2  e(\th )/T} \) \<\th^+ | \Phi |\th \>
\cr
& ~~~~~+ \inv{2} \int d1 d2 ~e^{- (e_1 + e_2 )/T }
\( 1 - e^{- e_1/T } - e^{- e_2 /T }  \)
\<2^+ ,1^+ |\Phi | 1,2\>  \cr
& ~~~~~+ \inv{3!}  \int d1 d2 d3 ~ e^{- (e_1 + e_2 + e_3 )/T}
      \<3^+ ,2^+ ,1^+ |\Phi |1,2,3\>
      +... )\cr }}
( $d1 = d\th_1 , e_1 = e(\th_1 ), {\rm etc.} $)
In the above formula, since we did the computation only up to 3
particles,
we only verified explicitly the appearance of $Z$ up to the
appropriate order, depending on which term $Z$ multiplies.

The above computation, along with some combinatoric checks at higher
order,
are sufficient to understand that to all orders one will find:
\eqn\eIIxiv{
\<\Phi\>_T =  \sum_{n=0}^\infty \inv{n!}
 \int {d\th_1 \cdots d\th_n \over (2\pi)^n}~ \prod_{i=1}^n  f(\th_i )
e^{- e(\th_i)/T }
       \< \th^+_n , \ldots , \th^+_1
       | \Phi  | \th_1 , \ldots , \th_n \>   , }
where
\eqn\eIIxv{
f(\th ) = \inv{ 1 + \exp( - e(\th) /T) } . }
In summary, the non-trivial effect of the disconnected pieces of the
form
factors is to modify the $\int d\th $ integrations by the factor
$f(\th )$.
In the thermodynamic approach, $f = \rho^h/( \rho + \rho^h )$.

The expression \eIIxiv\ has another kind of singularity arising from
poles in the connected piece of the form factor.  In fact, one of the
axioms of the form factor bootstrap expresses the residue of the pole
in $\<0| \Phi | ...\th, \th' - i\pi ,....\> $ as $\th \to \th'$ in
terms
of form factors with lower numbers of particles.
We will refer to these as pole singularities.  In the Ising model
with
$S=-1$, for the spin/disorder fields the residue axiom reads
\eqn\eres{
\< 0 | \Phi | \ldots \t , \t -i\pi + i\eta \ldots \> = \inv{ \eta}
\< 0 | \Phi | \ldots, \ldots \> . }
Thus,
\eqn\eresb{
       \< \th^+_n , \ldots , \th^+_1
       | \Phi  | \th_1 , \ldots , \th_n \>  =
\( \inv{ \eta} \)^n \< 0 | \Phi | 0 \> . }
This leads to
\eqn\eresc{
\< \Phi \>_T  =
\lim_{\eta \to 0}
\exp \(  \int {d\t\over 2\pi\eta} f(\t ) e^{-e(\t )/T } \)
\< \Phi \>_{T=0}
.}
Note that here we left $\eta$ inside the integration sign,
as it may well have to depend on $\beta$ when one tries to give a
more
precise meaning to \eresc (see the next appendix). Here, our attitude
will be to ignore this (infinite) multiplicative
renormalization, our goal being to obtain
expressions for the two point functions up to an overall
normalization, which
we set somewhat arbitrarily.

The above
procedure leads to meaningful expressions for the two-point
correlation
functions.  One begins with
\eqn\eIIxvii{
\< \Phi (x) \Phi (0) \>_T = \inv{Z} \sum_{\psi, \psi'}
e^{- E_\psi /T } {\<\psi| \Phi (x) | \psi'\> \<\psi' | \Phi (0) |
\psi \>
\over\<\psi|\psi\>\<\psi'|\psi'\>}. }
By explicitly evaluating the above sum for low numbers of particles,
one reaches the following conclusions.   Again, the disconnected
terms are of two types, one diverging in infinite volume but being
cancelled by $1/Z$, the other leading to modifications of the
integration
measures as for the 1-point function.   We find that the $\int d\th $
integrations for the sum over states $\psi$ and $\psi'$ are {\it
both}
modified.  To illustrate this, consider for instance the term:
\eqn\eIIxviii{
\inv{2} \int d\t d\t'_1 d\t'_2 e^{- e(\t )/T} \<\th | \Phi (x) |
\t'_1 ,
\t'_2 \> \<\t'_2 , \t'_1 | \Phi (0) |\th \> . }
One of the disconnected pieces is
\eqn\eIIxix{\eqalign{
- \frac{1}{2}  & \int
 d\t d\t'_1 d\t'_2 e^{- e(\t )/T}  \delta (\t -\t'_1 )
 \delta (\t - \t'_2 )  \<0 | \Phi (x) | \t'_2\>
 \<\t'_1 | \Phi (0) |0\>  \cr
 & ~~~~~ = - \frac{1}{2}
\int  d\t e^{- E(\t )/T}
   \<0 | \Phi (x) | \t\>
 \<\t | \Phi (0) |0\>  . \cr}}
This is interpreted as arising from the second term in the expansion
of
an $f(\th )$
measure factor in the sum over states $\psi'$.
In addition, the pole singularities can be removed by multiplicative
renormalization of the fields as described above.
 Based on this,
 one finds the  following expression for the 2-point functions
(again,
up to an overall normalization):
\eqn\eIIxx{\eqalign{
\<\Phi (x) \Phi (0) \>_T &=  \sum_{n,m = 0}^\infty
\inv{n! m! }  \int {d\t_1 ... d\t_n \over (2\pi)^n}
{d\t'_1 ....d\t'_m\over (2\pi)^m}
\( \prod_{i=1}^n f(\th_i ) e^{- e(\th_i )/T }  \)
\cr
& ~~~~~~\times \( \prod_{r=1}^m f(\th'_r )   \)\<0|\Phi (x) | \t'_1,
...,\t'_m , \t_n -i\pi, ...,\t_1
- i\pi \>
\cr
& ~~~~~~\times
      \<0| \Phi (0) | \t_1, ....,\t_n , \t'_m -i\pi, ....,\t'_1 -i\pi
\>
. \cr }}

\def\zbar{{\bar{z}}}

\def\ept{\tilde{\ep}}

Let us refer to the $\th'$ states in \eIIxx\ as particles and
the $\th$-states as holes.  Introducing an index $\ep = 1$ for
particles
and $\ep = -1$ for holes, define the particle-hole states
 as follows:
\eqn\estate{\eqalign{
 |\th_1 , \ldots, \th_n \>_{\ep_1\cdots \ep_n}  &=
 | \th_1 - \ept_1 i\pi  ,
\ldots , \th_n -  \ept_n i\pi \>  \cr
{}^{\ep_n \cdots \ep_1} \< \th_n, \ldots , \th_1 |
&=  \< \th_n +  \ept_n i\pi , \ldots , \th_1
+ \ept_1  i\pi |  , \cr }}
where $\ept = (\ep -1)/2 $.
The form-factors of the particle-hole states are the usual ones
but with the appropriate shifts by $i\pi$, e.g.
\eqn\ehole{
\< 0| \Phi (0) |\th_1 , \ldots, \th_n \>_{\ep_1\cdots \ep_n}  =
\< 0 | \Phi (0) | \th_1 - \ept_1  i\pi ,
\ldots , \th_n - \ept_n i\pi \> . }
The expansion \eIIxx\  can the be  written
as a sum over particles and holes:
\eqn\eIIxxi{\eqalign{
\<\Phi (x) \Phi (0) \>_T &=  \sum_{N= 0}^\infty
\inv{N! }  \sum_{\ep= \pm 1}
\int { d\t_1 ... d\t_N\over (2\pi)^N}
\[ \prod_{i=1}^N f_{\ep_i} (\th_i ) \exp \( - \ep_i (m z e^{\th_i}
+ m\zbar e^{-\th_i} )   \)   \]
\cr
& ~~~~~~\times \left|\<0|\Phi (0) | \t_1, ...,\t_N \>_{\ep_1 \cdots
\ep_N}\right|^2
, \cr }}
where
$f_\pm $ are defined in \filling.  This expression coincides with
\generes.

\appendix{B}{Derivation of $|B_T>$}

We work first at $T=0$. By considering $<0|{\cal O}|B>$, i.e.
in the L channel,
 and using the boundary state of \GZ,
we find the expression
\eqn\newi{\eqalign{<0|{\cal O}(x)|B>=\sum_n\int_{-\infty}^\infty
{d\beta_1\ldots
d\beta_{n}\over
 (4\pi)^{n} n!}
\rh (\beta_1)\ldots \rh (\beta_n)
<0|{\cal O}|-\beta_1,\beta_1,\ldots,-\beta_n,\beta_n>\cr
\exp\left[2mx(\cosh\beta_1+\ldots+\cosh\beta_n)\right].\cr}}

Now, suppose we want to compute this one point function in the
crossed, R channel. It then  reads simply $<0_B|\CO|0_B>$,
where $|0_B>$ is the ground state of the Ising model
on the half line with a magnetic field at $x=0$.

On the other hand,
let us perform crossing on \newi.
For this, let  us  first  set  $\beta\equiv i{\pi\over 2}-\beta'$.
 Using the
general expression
\eqn\crossing{<\beta_{n+1}\ldots\beta_{n+m}|{\cal
O}|\beta_1\ldots\beta_n>=<0|{\cal O}|
\beta_1,\ldots,\beta_n,\beta_{n+m}-i\pi,\ldots,\beta_{n+1}-i\pi>,}
one finds
\eqn\kki{\eqalign{
<0|{\cal O}|-\beta_1,\beta_1,\ldots,-\beta_n,\beta_n>
=(-1)^n<-\beta'_n,\ldots,-\beta'_1
|{\cal O}|\beta'_1,\ldots,\beta'_n>\cr
=<-\beta'_n+i\pi,\ldots,-\beta'_1+i\pi|{\cal O}|\beta'_1-i\pi,
\ldots,\beta'_n-i\pi>.\cr}}
The $\beta'$ integration runs now from ${i\pi\over 2}-\infty$ to
${i\pi\over 2}+\infty$. Let us move  this contour  back to the real
axis - to do so, we request
that $\rh$ has no pole in the physical strip, which will
be the case for appropriate magnetic field \GZ. Using  that
$\rh (\beta)=R\left({i\pi\over 2}-\beta\right)$, together with
$R(\beta)=-[R(\beta-i\pi)]^*$  the one point function \newi\
can be then rewritten as
\eqn\newii{\eqalign{\sum_n\int_{-\infty}^\infty {d\beta_1\ldots
d\beta_{n}\over
(4\pi)^{n} n!}(-1)^n\left[R(\beta_1-i\pi)\ldots
R(\beta_n-i\pi)\right]^*\cr
<-\beta_n+i\pi,\ldots, -\beta_1+i\pi
|{\cal O}|\beta_1-i\pi,\ldots,\beta_n-i\pi> \cr
\exp\left\{-2mix[\sinh\beta_1+\ldots+\sinh\beta_n]\right\}.\cr}}

Expression \newii\ can be explained by realizing that the  boundary
changes the nature of the Fermi sea of the  quantum theory. From
\newii, we
see that we can represent the new ground state  $|0_B>$  by
writing  formally  as $|0_B>=\otimes_{\beta>0}\left\{
|\beta-i\pi>-R(\beta-i\pi)|-\beta-i\pi>\right\}$. Indeed, evaluate
 now the one point function of ${\cal O}$
 on the half line , $<0_B|{\cal O}|0_B>$.
We first have two terms where no $R$ matrix is involved,
and which reads naively, using crossing,
\eqn\junk{
\left(\otimes_{\beta>0} <\beta|\right){\cal
O}(x)\left(\otimes_{\beta>0}|\beta>\right)=\prod
{1\over \eta(\beta)},}
together with a similar expression with $\beta\to -\beta$. Here
$\eta(\beta)$ is a cut-off  necessary  to avoid the
divergences due to particle-antiparticle annihilation. The general
form
of $\eta$ is easy to obtain:
at finite   $L$, require that a particle and an antiparticle cannot
occupy the same state: this leads to a
a cut-off in rapidity space such that
$\delta p(\beta)=m\cosh\beta\delta\beta={\pi\over L}$,
or $\delta(\beta)={\pi\over mL\cosh\beta}$ \foot{Here
we used the fact that with a boundary, while only positive
rapidities are allowed, the density is twice as big}. We use this
cut-off as
the
value of $\eta(\beta)$ in what follows.

A renormalization of the operator ${\cal O}$ of course
suppresses the term on the right hand side of \junk. Let us
set  $\left(\otimes_{\beta>0} <\beta|\right){\cal
O}(x)\left(\otimes_{\beta>0}
 |\beta>\right)+(\beta\to -\beta)=1$ (much
the same renormalization would occur in the bulk). Note that there
is no $x$ dependence so far.

 Then, there are also crossed
terms
in $<0_B|{\cal O}|0_B>$, which
involve the matrix $R$. These are $x$ dependent and cannot be
absorbed
in a renormalization.
Indeed, for
 a given rapidity $\beta$ in the decomposition of the ket $|0_B>$,
in the corresponding bra  we can pick either $\beta$ or $-\beta$.
In that case, compared with  \junk\  and its $\beta\to -\beta$
equivalent we get a relative factor
\eqn\junki{{<-\beta|{\cal O}|\beta> \over
<\beta^+|{\cal O}|\beta>+<-\beta^+|{\cal O}|-\beta>}={<-\beta|{\cal
O}|\beta>\over
2{mL\cosh\beta\over
\pi}}.}
On the other hand, this particular rapidity where a pairing
with a reflected state takes place can be chosen with some
multiplicity: there
are ${L\over \pi}\cosh\beta d\beta$ states with which one can do
so
in $[\beta,\beta+d\beta]$
Clearly, the two
L-dependent  contributions
cancel out, leaving an overall factor of $1/2$. Generalizing to
arbitrary number of
 pairings
and using the fact that $R^*(\beta)=R(-\beta)$, one
reproduces
the expression \newii, where the integrals with $\beta>0$ corresponds
to taking the reflected state in the bra, the integral with $\beta<0$
to taking the reflected state in the ket.

The foregoing form of $|0_B>$
is easily checked (eg by evaluating matrix elements
between $|0_B>$ and any other state) to satisfy the relations \GZ\
 $A^\dagger(\beta)|0_B>=R(\beta)A^\dagger (-\beta)|0_B>$, using that
 $-[R(\beta-i\pi)]
={1\over R(\beta)}$.

To  extend  the result at finite $T$, we require similarly for an
excited
state $|\beta_B>$ to satisfy
$A(\beta)|\beta_B>=R^*(\beta)A(-\beta)|\beta_B>$.
Using $R^*(\beta)={1\over R(\beta)}$, it follows that an excited
state
is obtained by replacing some of the
$|\beta-i\pi>-R(\beta-i\pi)|-\beta+i\pi>$
by $|\beta>+R(\beta)|-\beta>$. This leads to two types of states in
the boundary heat
bath $|0_{B,T}>$ characterized by a label $\epsilon=\pm 1$, with
relative weights $f_{\pm}(\beta)$. As a result, the one point
function
at finite temperature reads then

\eqn\newiii{\eqalign{<0_{B,T}|{\cal O}(x)|0_{B,T}>=&
\sum_n\int_{-\infty}^\infty {d\beta_1\ldots
d\beta_n\over(4\pi)^n n!}\sum_{\epsilon_i} (-1)^{\bar{\epsilon}_i}
\left[R(\beta_1-\bar{\epsilon}_1i\pi)\ldots
R(\beta_n-\bar{\epsilon}_n i\pi)\right]^*\cr
&<-\beta_n+\bar{\epsilon}_n i\pi,\ldots,
-\beta_1+\bar{\epsilon}_1i\pi
|{\cal
O}|\beta_1-\ept_1i\pi,\ldots,\beta_n-\ept_ni\pi>
\cr
&\prod_i\exp[2mix\epsilon_i\sinh\beta_i]
f_{\epsilon_i}(\beta_i),\cr}}
where we set $\bar{\epsilon}\equiv{1+\epsilon\over 2}$.

One can then get back to the original channel by following the
previous transformations in the opposite way to find \bconj,\tough.

\appendix{C}{Boundary Field Theory on a Finite  Strip}


The techniques of this paper  allows us to consider
a different geometry, that of an infinitely long strip of finite
width $R$.  In this geometry one can consider a quantum field
theory with non-trivial boundary interactions at $x=0,R$,
where the strip is defined by  $0\leq x \leq R$.
Here, $R$ does not have the interpretation as an inverse temperature.
Let us quantize the theory in a way that
views $x$ as the imaginary time.  In this quantization the
Hamiltonian $H$
is the usual bulk hamiltonian with no boundary terms, and the
spectrum
and form factors are the same as for a bulk theory with hamiltonian
$H$.  In this picture, the boundary interactions are contained in
a boundary state as in \rtti.

Consider the one-point function of a field $\Phi (x)$.
For $x$ a semi-infinite line with $R=\infty$,
this correlator was characterized with form factors
in \rkon.
Due to the
translation invariance in the $t$-direction, this correlator depends
only on $x$.  For the finite width strip, one is allowed to put
different
boundary conditions at each end of the strip.  Let the boundary
condition at $x=0$ correspond to the boundary state $|B_b \>$,
and at $x=R$ the boundary state $|B_a\>$.  The partition function
in this picture is then
\eqn\epartab{
Z_{ab} (R)  = \< B_a | e^{-RH} |B_b \> . }
The one-point function
is defined  by
\eqn\eIVii{
\<B_a |  \CO (x) |B_b \>_R = \inv{Z_{ab}(R)}
\< B_a | e^{-H(R-x)} \CO (0) e^{-Hx} |B_b \> . }

Since the boundary state involves pairs of particles with opposite
rapidity, the states which contribute as intermediate states  are
\eqn\eIViii{
| 2n \> = | -\t_1 , \t_1 , \ldots , -\t_n ,  \t_n \> . }

Inserting two complete sets of states in \eIVii, one on each side
of the field $\CO$, one obtains
\eqn\eIVvb{
\eqalign{
\< B_a | \Phi (x) |B_b \>_{R} =  \sum_{n,m=0}^\infty
\inv{n! \, m! }
\int_{-\infty}^\infty {d\t_1 \cdots d\t_n \over (4\pi)^n}
\int_{-\infty}^\infty {d\t'_1 \cdots d\t'_m\over (4\pi)^m} ~\cr
\times\( \prod_{i=1}^n e^{-2 (R-x) e(\t_i )  } \rh^*_a (\t_i ) \)
  \( \prod_{i=1}^m  e^{-2x e(\t'_i )} \rh_b (\t'_i ) \)\cr
\times\< \t_n , -\t_n , \ldots ,  \t_1 , -\t_1 | ~ \Phi (0) ~
| -\t'_1 , \t'_1 , \ldots , -\t'_m , \t'_m \> , \cr }}
where as usual $e(\t ) = m \cosh \t $.

As in appendix A, one finds that the disconnected pieces of the form
factors can be incorporated by simply modifying the measures for the
integrals over $\th$.  Also as before the residue singularities
lead to a multiplicative renormalization of the fields, which we
again
ignore.
It is not hard to see that one obtains
\eqn\eIVviii{\eqalign{
\< B_a | \CO (x) |B_b \>_{R}
=  \sum_{n,m=0}^\infty
\inv{n! \, m! }
\int_{-\infty}^\infty {d\t_1 \cdots d\t_n \over (4\pi)^n}
\int_{-\infty}^\infty {d\t'_1 \cdots d\t'_m\over (4\pi)^m}\cr
\times\( \prod_{i=1}^n e^{- 2 (R-x)  e(\t_i )  } \rh^*_a  (\t_i ) f
(\t_i ) \)
\( \prod_{i=1}^m  e^{-2 x e(\t'_i ) } \rh_b (\t'_i )  f (\t'_i ) \)
\cr
\times\< \t^+_n , -\t^+_n , \ldots ,  \t^+_1 , -\t^+_1 | ~ \Phi (0) ~
| -\t'_1 , \t'_1 , \ldots , -\t'_m , \t'_m \> , \cr }}
where now
\eqn\eIVix{
f (\t ) = \inv{ 1 + e^{-2 Re (\t ) } \rh_b(\t) \rh^*_a  (\t ) } . }

Result \eIVviii\ can be put in a thermodynamic
form by defining $f_+(\beta)=f(\beta)$ as in \eIVix, and
$f_-(\beta)=1-f(\beta)$.
Then, using that $\hat{R}(\beta-i\pi)=-{1\over \hat{R}(\beta)}$,
one finds
%
\eqn\newtherm{\eqalign{<B_a|\CO(x)|B_b>_R=\sum_{N=0}^\infty\sum_{
\epsilon=\pm}
{1\over N!}\int_{-\infty}^\infty\prod_{i=1}^N \left\{{d\beta_i\over
4\pi}e^{-2x\epsilon_ie(\beta_i)} [\hat{R}_b]_\epsilon(\beta_i)
f_\epsilon(\beta_i\right\}\cr
 <0|\CO|-\beta_1,\beta_1,\ldots,-\beta_N,\beta_N>_{\epsilon_1,\ldots,
\epsilon_N}.\cr}}
This can easily be recovered in a thermodynamic approach.
 The TBA was developed for the partition functions
$Z_{ab}$ in \ref\rbtba{A. Leclair, G. Mussardo, H. Saleur, and S.
Skorik,
Nucl. Phys. B {\bf 453} (1995) 581.},
and one sees that $f$ given in \eIVix\ corresponds precisely to the
appropriate ratio of densities
as determined there.

\appendix{D}{Computations for bulk correlators}
In this appendix we shall provide the missing steps required to
establish \essnii.

First, we show that the analog of \essni\ with $T$ derivatives also
holds.
As in \essIIi\ we have
\eqn\edi{T \partial_T \ln \tau_{\pm} = \pm \Tr \(
( 1- \R_{\pm} ) T \partial_T \K \) .}
 From the defining relations \eIIIxpp\ and \eIIIxb\ we have
\eqn\edii{T \partial_T \K (u,v) = \frac{T \partial_T f (u)}{2 f(u)}
\K (u,v) + \K (u,v)  \frac{T \partial_T f (v)}{2 f(v)}.}
Inserting this into \essIIi, and using \eIIIxi\ we get
\eqn\ediii{T \partial_T \ln \tau_{\pm} =  \Tr \( \R_{\pm} (u,u)
\frac{T \partial_T f (u)}{ f(u)} \)}
where the symbol $\Tr$ implies an integral from $-\infty$ to $\infty$
over variables
$u, v, \ldots$ that appear in its argument. Now using the explicit
expression for
$\R$ in \eIIIxvii\ and \eIIIxxvii\ we get
\eqn\ediv{T \partial_T \ln \tau_{\pm} =  \frac{1}{2} \Tr \( \(
\partial_u g_- (u) g_+ (u) - g_- (u) \partial_u g_+ (u) \pm
\frac{g_- (u) g_+ (u)}{u} \)
\frac{T \partial_T f (u)}{ f(u)} \). }
In particular, we have
\eqn\edv{T \partial_T \ln ( \tau_- / \tau_+ ) =   - \Tr \(
\frac{g_- (u)}{u}
\frac{T \partial_T f (u)}{ f(u)} g_+ (u) \). }
In the following, we will show that the right hand side of \edv\ is
simply related
to the potentials which were introduced earlier. We have also
examined
$T \partial_T \ln ( \tau_-  \tau_+ )$, but its value does not appear
to be
expressible solely in terms of the potentials of Section 4.3.

Apply the operator $( 1 - \R_{\pm} ) T \partial_T $ to both sides of
the first
equation in \eIIIxxva. Using \edii\ and simplifying we get
\eqn\edvi{ T \partial_T g_{\pm} (u) =
 \frac{T \partial_T f (u)}{2 f(u)} g_{\pm} (u) -
\int_{-\infty}^{\infty} dv
\R_{\pm} (u, v)
\frac{T \partial_T f (v)}{ f(v)} g_{\pm} (v).}
Also, from \eIIIxpp\ we have
\eqn\edvii{ T \partial_T e (u) =
 \frac{T \partial_T f (u)}{2 f(u)} e (u).}
We are now ready to take $T$ derivatives of the potentials. We have
from the
definition \essIIvii\ and \edvi, \edvii
\eqn\edviii{
\eqalign{ T \partial_T c_{\pm} & = T \partial_T \Tr ( e \d_{-1}
g_{\pm} ) \cr
& = \Tr \left( e \d_{-1} ( 1 - \R_{\pm} )  \frac{T \partial_T f }{ f}
g_{\pm}
\right)  \cr
& = ( 1 \mp c_{\pm} ) \Tr \left( g_{\mp} \d_{-1} \frac{T \partial_T
f }{ f} g_{\pm}
\right), \cr}}
where in the last step we have used the intertwining relations
\essIiib.
Finally, using the parametrization \essIIviiaa, and comparing with
\edv, we get
\eqn\edix{T \partial_T \left( \ln (\tau_{-}/\tau_{+}) - \phi \right)
= 0.}

The combination of \essni\ and \edviii\ implies that $\tau_- / \tau_+
= C e^{\phi}$ with $C$ a pure number independent of $z$, $\zbar$, and
$T$.
We now determine $C$ by computing the large $z$, $\zbar$ behavior of
$\tau_{\pm}$ and $\phi$ at $T=0$ directly from the form factor
expansion.
Indeed, simply by evaluating the first two terms in the summation in
\eIIIvi\ we have
\eqn\edx{
\tau_{\pm} = 1 \pm \frac{e^{-mr}}{\sqrt{2 \pi m r}} + \ldots
{}~~~~~~~T=0,~r \rightarrow \infty}
where $r^2 = 4 z \zbar = x^2 + t^2$. Inserting \edx\ into the first
differential equation in \etaub\ we get
\eqn\edxi{
\phi = - \frac{2 e^{-mr}}{\sqrt{2 \pi m r}} + \ldots
{}~~~~~~~T=0,~r \rightarrow \infty}
Finally, inserting \edx, \edxi\ into  $\tau_- / \tau_+
= C e^{\phi}$ and matching the large $r$ behavior, we get $C=1$.

\listrefs

\bye